\RequirePackage{etex} 
\documentclass[aps,prb,english,longbibliography,twocolumn,superscriptaddress]{revtex4-2}
\usepackage[active]{srcltx}
\usepackage[utf8]{inputenc}
\usepackage{latexsym}
\usepackage[colorlinks=true,linkcolor=blue,citecolor=red]{hyperref} %
\usepackage{orcidlink}
\usepackage{amsmath}
\usepackage{xcolor}
\usepackage{amssymb}
\usepackage{amsfonts}
\usepackage{graphicx}
\usepackage{slashed}
\usepackage{wasysym}
\usepackage{pifont}
\usepackage{manfnt}
\usepackage{tikz}
\usepackage{tikz-3dplot}
\usepackage{tikz-network}
\DeclareSymbolFont{usualmathcal}{OMS}{cmsy}{m}{n}
\DeclareSymbolFontAlphabet{\mathcal}{usualmathcal}
\usepackage{braket}
\usepackage{subfig}
\usepackage{ragged2e}

\newcommand{\be}{\begin{equation}}
	\newcommand{\ee}{\end{equation}}
\newcommand{\bea}{\begin{eqnarray}}
	\newcommand{\eea}{\end{eqnarray}}

\newcommand{\mc}{\mathcal}

\begin{document}
\title{Symmetry protected quantum many body scars through half-gauging}
\author{Weslei B. Fontana \orcidlink{0000-0001-8368-5997}}
\email[Email adress:]{weslei.bernardinofontana@oist.jp}
\affiliation{Theory of Quantum Matter Unit, Okinawa Institute of Science and Technology Graduate University, Onna-son, Okinawa 904-0412, Japan}
\author{Fabrizio G. Oliviero \orcidlink{0000-0002-5721-7687}}
\email[Email addres:]{fabrizio@phys.ncts.ntu.edu.tw}
\affiliation{Department of physics, National Tsing Hua University, Hsinchu 30013, Taiwan}
\affiliation{Center for Theory-Computation-Data Science Research (CTCD), 5th Floor, 3rd General Building, National Tsing Hua University, No. 101, Sec. 2, Kuang-Fu Rd., Hsinchu, 300044, Taiwan}
\affiliation{Physics Division, National Center for Theoretical Sciences, Taipei 10617, Taiwan}
\author{Yi-Ping Huang\,\orcidlink{0000-0001-6453-1191}}
\email[Email address:]{yphuang@phys.nthu.edu.tw} 
\affiliation{Department of physics, National Tsing Hua University, Hsinchu 30013, Taiwan}
\affiliation{Center for Theory-Computation-Data Science Research (CTCD), 5th Floor, 3rd General Building, National Tsing Hua University, No. 101, Sec. 2, Kuang-Fu Rd., Hsinchu, 300044, Taiwan}
\affiliation{Physics Division, National Center for Theoretical Sciences, Taipei 10617, Taiwan}
\affiliation{Institute of Physics, Academia Sinica, Taipei 115, Taiwan}
\begin{abstract}
We study the interplay between symmetry defects insertion and the phenomena of quantum many body scarring. We exploit a class of models obtained from an embedding mechanism that defines a \textit{stochastic matrix form} hamiltonian, hosting zero energy many body scarred states. Such models display an interesting playground to explore the consequences of dualities, half-gauging operations and so on. We first show that inside this class of models, what are the conditions to still have scared states after defect insertion. The main result of our work is a detailed analysis of half-gauging in such chains, which introduces a non-trivial interface in our system, leading to scarred states protected by the underlying symmetries. The conclusions for half-gauged scarred systems are general and may lead to fruitful directions for higher dimension scarred physics and their robustness. 
\end{abstract}

\maketitle

\section{Introduction}

Quantum many-body systems at finite energy density are generally expected to
thermalize according to the Eigenstate Thermalization Hypothesis (ETH)
\cite{VonNeumann2010,deutsch1991,Srednicki1994,DAlessio2016}.
Quantum many-body scars (QMBS) provide a striking exception to this
expectation: they form atypical nonthermal eigenstates embedded within an
otherwise thermal many-body spectrum
\cite{Turner2018,Papic2018,Serbyn2021,Moudgalya2022,
chandran_quantum_2023}.
Following their experimental observation in synthetic quantum systems
\cite{Bernien2017}, substantial progress has been made in understanding how
such states arise and how they can be embedded exactly into otherwise
thermal spectra
\cite{odea2020,Odea2024,Shiraishi2017,Moudgalya2024,Motrunich2020,
TaoLin_2025,Nicolau2025,Budde2026}.
These developments naturally raise a broader question beyond the existence
of individual scarred eigenstates: \emph{what structures of quantum states
and phases remain meaningful when they are embedded into a nonthermal
subspace at finite energy density?}

Ground-state quantum phases provide a natural setting in which to address
this question. Symmetry-protected topological (SPT) and symmetry-breaking
phases are distinguished not merely by particular ground-state wave
functions, but by how their states respond to symmetry defects, boundaries,
and gauging. Interfaces generated by such operations can furthermore carry
nontrivial symmetry algebras and protected degeneracies. States belonging
to SPT and symmetry-breaking phases can also serve as anchor states for
embedding exact QMBS \cite{Chamon2019}. Embedding therefore poses a basic
conceptual question: does it merely relocate particular wave functions into
the middle of the spectrum, or can the organizing structures associated
with their quantum phases survive at finite energy density? More
specifically, can both the distinction between SPT and ordered states and
the nontrivial structure associated with their symmetry interfaces be
carried into an embedded nonthermal subspace?

Gauging and non-invertible duality provide a controlled way to formulate
these questions. Recent works have shown that such transformations can be
compatible with QMBS embedding mechanisms and can generate new scarred
Hamiltonians \cite{Katsura2025,Fontana2025}. In our previous work, we showed
that the survival of QMBS under Kramers--Wannier duality is controlled by
whether the duality transformation preserves the underlying embedding
condition \cite{Fontana2025}. That result established when duality is
compatible with a scar embedding. Here we ask a different question:
\emph{whether the phase and interface structures familiar from ground-state
physics themselves remain consequential after embedding.} The $k$-chain
family provides a particularly useful setting for answering this question,
because its anchor states span both SPT and symmetry-breaking phases and
admit symmetry-defect and gauging constructions that can be followed
explicitly after embedding
\cite{Suzuki1971,Verresen2017,Tantivasadakarn2023,Chamon2019}.

Our first result shows that embedding does not erase the distinction between
the underlying phases. We construct scarred $k$-chains from SPT and
symmetry-breaking anchor states and examine their response to symmetry
defects. For SPT anchors, the complete QMBS subspace survives the
corresponding defect insertion and is mapped by gauging onto a complete dual
scarred subspace. Under open boundary conditions (OBC), however, the
embedding preserves the degeneracy of the SPT anchor subspace while the
original edge-mode operators cease to commute with the scarred Hamiltonian.
The resulting QMBS degeneracy is therefore no longer enforced by these
exact boundary zero modes, revealing a distinction between preserving the
anchor subspace and preserving its edge-mode protection.
For symmetry-breaking anchors, by contrast, defect insertion obstructs the
embedding condition, and only the symmetry-compatible sector survives
gauging. Thus, the SPT-versus-ordered character of the anchor state remains
physically consequential after embedding: it determines the response of
the finite-energy nonthermal subspace to symmetry defects and gauging.
In this sense, a structural distinction conventionally associated with
ground-state phases is inherited by the embedded QMBS subspace.

This result motivates a stronger question. In ground-state physics, the
structure associated with a quantum phase is not exhausted by the bulk
state itself. Symmetry defects and gauging can generate interfaces carrying
nontrivial symmetry actions and protected degeneracies. We therefore ask
whether such an interface structure can itself be embedded at finite energy
density while remaining compatible with the exact QMBS construction. This
compatibility is not automatic: introducing an interface modifies the
Hamiltonian and can in principle destroy the conditions responsible for
the scar embedding. If the two structures can coexist, however, embedding
would preserve not only information about the underlying phase, but also
the protected algebraic structure associated with its interface.

We show that this compatibility is realized through \emph{half-gauging}.
Gauging the symmetry over only part of the system creates an interface
between the original scarred Hamiltonian and its gauged dual while
preserving the QMBS embedding. The symmetry operators associated with the
two sides of the interface realize a nontrivial anomalous algebra. Its
multidimensional irreducible representations enforce additional spectral
degeneracies and consequently enlarge the embedded QMBS subspace into a
symmetry-protected multiplet. We refer to these interface-enriched scarred
subspaces as \emph{anomalous QMBS subspaces}, reflecting the anomalous
symmetry algebra acting within them. We demonstrate this mechanism
explicitly for several $k$-chains, where the enlargement of the scarred
subspace is determined by the representation structure of the interface
symmetry algebra. The resulting degeneracy is protected against
perturbations that preserve the relevant symmetry algebra, closely
paralleling the spectral constraints associated with strong zero modes
\cite{Aasen2016,Fendley2016,Moudgalya2026}. In particular, under OBC, half-gauging
restores a protected-mode interpretation of the QMBS degeneracy through
nonlocal conserved operators extending from the interface.

Our results therefore establish two complementary connections between
ground-state phase structure and embedded nonthermal subspaces. First, the
SPT-versus-ordered distinction of the anchor state survives embedding
through the characteristic response of the QMBS subspace to symmetry
defects. Second, the anomalous structure associated with a symmetry
interface can itself coexist with the embedding and endow the QMBS subspace
with additional protected degeneracy. Embedding thus need not erase the
organizing structures of quantum phases: both bulk phase information and
interface-induced algebraic structure can remain operative within
finite-energy nonthermal subspaces. This perspective moves beyond the
construction of individual scarred eigenstates toward the engineering of
structured nonthermal subspaces, and suggests a route for carrying
symmetry- and interface-based organizing principles into scarred systems in
higher dimensions.

This paper is organized as follows. In Sec.~\ref{secII}, we introduce the
$k$-chains and construct the corresponding family of scarred Hamiltonians.
In Sec.~\ref{secIII}, we study their response to symmetry-defect insertion
and establish how the SPT or symmetry-breaking character of the anchor state
remains encoded in the resulting QMBS subspace. In Sec.~\ref{secIV}, we
construct half-gauged scarred systems and show how the resulting interfaces
generate anomalous symmetry algebras and protected degeneracies within the
embedded subspace. We conclude in Sec.~\ref{secV} with a summary and
discussion of future directions.

\section{$k$-chains and embedding structures \label{secII}}
The class of models we are interested are commonly referred to as $\alpha$-chains. For simplicity of the upcoming notation, we will use the nomenclature of $k$-chains in this work, in similarity with definitions used in \cite{Tantivasadakarn2023}. This class of models first appeared in the literature in the 1970s \cite{Suzuki1971} in the context of dualities involving classical dimer models. Since then, they have attracted renewed interest, primarily due to their exotic critical properties \cite{Ville2015, Verresen2017}.

These models can be compactly defined in terms of Pauli operators acting on strings supported over $k$ sites
\begin{equation}
H^{(k)} = \begin{cases}
-\sum_j Y_jX_{j+1}\dots X_{j+k-1}Y_{j+k}\,, ~ ~ ~ &\text{if $k<0$}\,,\\
-\sum_j X_j\,,~ ~ ~ &\text{if $k=0$}\,,\\
 -\sum_j Z_{j}X_{j+1}\dots X_{j+k-1}Z_{j+k}\,,~ ~ ~ &\text{if $k>0$}\,.
   \end{cases}\label{eq:1}
\end{equation}

For $k>0$, we can already identify some known standard models, such as the classical Ising for $k=1$, the cluster model (or ZXZ model) for $k=2$ and so forth. Generically, these models have a $\mathbb{Z}_2^k$ symmetry defined by 
\begin{equation}
    \eta^{(a)} = \prod_{i\in \Lambda_a} X_i\,,~~~a=0,\,\ldots,\,k\,.
    \label{eq:z2ksymmetries}
\end{equation}
where $\Lambda_k$ labels each of the possible $k$ sublattices; for example, $k=2$ we have the standard cluster model with its $\mathbb{Z}_2\times\mathbb{Z}_2$ symmetry.

For notational convenience, we write the Hamiltonians in \eqref{eq:1} in the generic form $H^{(k)} = -\sum_j C^{(k)}_j$, where $C^{(k)}$ denotes the product of Pauli matrices acting on $k$-sites as depicted in \eqref{eq:1}. 

The ground state physics of such chains has been previously explored in \cite{Verresen2017}. For $k\in\mathbb{Z}_{\text{odd}}$, the ground state corresponds to a spontaneous symmetry breaking phase with respect to the global $\mathbb{Z}_2$ symmetry. Indeed, it is straightforward to show that $\prod_j C^{(2k+1)}_j\sim \openone$ while $\prod_j C^{(2k)}_j\sim \prod_j X_j$. Since $(C^{(k)})^2=\openone$ for any $k$, this implies that for $k\in \mathbb{Z}_{\text{odd}}$, these ground states are 2-fold degenerate with the global $\mathbb{Z}_2$ relating the two ground states. Conversely, for $k\in\mathbb{Z}_{\text{even}}$, the ground state typically lies within some symmetry protected topological (SPT) phase \cite{Verresen2017, Tantivasadakarn2023}. 

Following the procedure \cite{Chamon2019}, we can define a class of scarred $k$-chains models via the deformation
\begin{equation}
H^{(k)}_s = \sum_j \alpha_j\left(e^{-\beta\left(X_j+X_{j+k}\right)} - C^{(k)}_j\right)\,.
    \label{eq:scarred models}
\end{equation}
For an appropriate choice of the $\alpha_j$ coefficients \cite{Chamon2019, Fontana2025}, this Hamiltonian hosts mid-spectrum zero energy states of the form
\begin{equation}
    \ket{C^{(k)}(\beta)} = e^{\frac{\beta}{2}\sum_jX_j}\ket{C^{(k)}}\,,
\end{equation}
where $C^{(k)}_j\ket{C^{(k)}} = \ket{C^{(k)}}$ for all $j$. These states exhibit area-law entanglement, as they are generated from the action of a low bond dimension matrix product operator (MPO) on top of a stabilizer state. 

Recognizing that the $X$-Pauli matrix in the exponential corresponds to the $0$-chain, we can systematically map out a whole web of dualities involving these scarred models using the pivot framework \cite{Tantivasadakarn2023}. For instance, while the standard Kramers-Wannier duality acts as $C^{(k)}_j\mapsto C^{(1-k)}_j$, pivoting along a $k_0$-chain via the unitary transformation
\begin{equation}
    U_{k_0} = \exp\left(\frac{i\pi}{4} H^{(k_0)}\right)\,,
\end{equation}
yields the mapping $H^{(k)}\mapsto H^{2k_0 -k }$. Applying this map to the scarred Hamiltonian \eqref{eq:scarred models} yields another scarred Hamiltonian with the form
\begin{equation}
    H^{(k)}_s\mapsto \sum_j\alpha_j\left(e^{-\beta\left(C^{(2k_0)}_j+C^{(2k_0)}_{j+(2k_0-k)}\right)} - C^{(2k_0 - k)}_j\right)\,,
\end{equation}
where the zero-energy states are given by
\begin{equation}
    \ket{C^{(2k_0-k)}(\beta)} = e^{\frac{\beta}{2}\sum_jC^{2k_0}}\ket{C^{(2k_0 -k)}}\,.
\end{equation}

While we have established that these states exhibit low entanglement and lie in the middle of the spectrum, it remains to be shown that these scarred models are non-integrable. This can be demonstrated constructively. Specifically, the $1$-chain model was shown to be non-integrable based on its Wigner-Dyson level spacing distribution \cite{Chamon2019}. 

A similar argument to the one presented in \cite{Chamon2019}, also applies here. The idea is to explore an identification between the spectrum of the $k$-chains with the one of the $1$-chain. Since the global $\mathbb{Z}_2^{k}$ symmetries \eqref{eq:z2ksymmetries} act on $k$ distinct sublattices, we can decompose the scarred Hamiltonian into a sum of terms restricted to each of those sublattices:
\begin{equation}
    H^{(k)}_s = H^{(k)}_1+\dots+H^{(k)}_k\,.
\end{equation}

In the above, the superscript in parentheses labels the parent $k$-chain Hamiltonian, while the subscript indicates the sublattice on which the terms act. Because these terms have disjoint support, they trivially commute $\left[H_i,\,H_j\right]=0$.

Furthermore, for $H_{i\in \Lambda_a}$ and $X_{i\in \Lambda_b}$ we also have $\left[H_{i\in\Lambda_{a}},\, X_{i\in\Lambda_b}\right]=0$ whenever $a\neq b$. This implies that each of the $X_i$ operators act as local symmetries for the individual Hamiltonian terms on different sublattices. Consequently, we can analyze each of the $H^{(k)}_{i\in \Lambda_a}$ individually fixing the eigenvalues of $X_{i\notin\Lambda_a}$ as conserved charges. This projection reveals that the spectrum of $H^{(k)}_i$, restricted to the $X_i=+1$ charge sector, is equivalent to that of the scarred $1$-chain, which has already been shown to obey Wigner-Dyson level spacing statistics \cite{Chamon2019, Fontana2025}. We therefore conclude that the generalized scarred models \eqref{eq:scarred models} indeed hosts well-defined QMBS states in the middle of their spectra.
\section{Warmup: Defect insertion and conservation of the embedding condition \label{secIII}}
In this section we focus on a few simple examples, specifically the $2$- and $3$-scarred chains, and analyze sectors of the Hilbert space that contain a symmetry defect. The case of the $1$-scarred chain has already been discussed in \cite{Fontana2025}. Our goal is to demonstrate that for certain $k$-scarred chains, defect insertion does not spoil the embedding condition of the models in \eqref{eq:scarred models}, thereby preserving the existence of the scarred states in these sectors. 

This analysis is crucial for the discussion of duality relations among these models, as the scrambling of the Hilbert space under a duality transformation may or may not preserve the scarred states. Specifically, we aim to identify the necessary conditions under which a given duality guarantees the existence of a corresponding dual scarred state, at least for the class of models described by \eqref{eq:scarred models}.

\subsection{$2$-chain: Cluster model embedding}
The $2$-chain is equivalent to the standard cluster model with $\mathbb{Z}_2\times\mathbb{Z}_2$ symmetry, defined by 

\begin{equation}
    H^{(2)} = -\sum_j Z_jX_{j+1}Z_{j+2}\,.
\end{equation}

The corresponding scarred model is obtained by setting $k=2$ into \eqref{eq:scarred models}, which leave us with
\begin{equation}
    H^{(2)}_s = \sum_{a,\,j}\alpha_{j,\,a}\left(e^{-\beta_a\left(X_j+X_{j+2}\right)} - Z_jX_{j+1}Z_{j+2}\right)\,.
    \label{eq: alpha 3 scar hamilt}
\end{equation}
The index $a$ specifies the sublattice on which the alternating pattern is
defined. For $a=0$, $j$ takes even values, while for $a=1$, $j$ takes odd
values. For $a=0$, $$\alpha_j^{(0)}=|\alpha|+(-1)^{j/2},\qquad j=0,2,4,\ldots$$ Similarly, for $a=1$,$$
\alpha_j^{(1)}=|\alpha|+(-1)^{(j-1)/2},\qquad j=1,3,5,\ldots$$
This particular deformation was previously studied in \cite{Chamon2019}. To lay the groundwork for analyzing the dualities of this model, we show that the scar embedding conditions remain satisfied even in the presence of symmetry defects.

We consider a closed chain of $N$ sites. The $\mathbb{Z}_2\times \mathbb{Z}_2$ global symmetry of the model is generated by the product of $X$ Pauli operators on the respective sublattices:
\begin{equation}
    U^e=\prod_{i\in \Lambda_e}X_i\,, ~ ~ ~ U^o=\prod_{i\in \Lambda_o} X_i\,,
\end{equation}
where $\Lambda_e,\,\Lambda_o$ denote the even and odd sublattices, with eigenvalues $(-1)^{u_e}$ and $(-1)^{u_o}$, where $u_e,\,u_o\in \left\{0,\,1\right\}$, respectively. Moreover, it is also convenient to consider twisted sectors in the Hilbert space, labeling the full Hilbert space tuple $(u_e,\,u_o,\,t_e,\,t_o)$, where $u_{e,\,o}$ specify the symmetry sectors and $(t_e,\,t_o)\in\left\{0,\,1\right\}$ specify the twist sectors, with eigenvalues $(-1)^{t_e}$ and $(-1)^{t_o}$, respectively. Being in a twisted sector is equivalent to the insertion of a symmetry defect for the corresponding symmetry generator.

We now proceed to show, that embedding conditions for the QMBS of the cluster model remain preserved in twisted sectors. To start, consider the Hamiltonian \eqref{eq: alpha 3 scar hamilt} in untwisted sectors $(u_e,\,u_o,\,0,\,0)$. The QMBS state correspond to a cluster state acted upon by a local operator. Explicitly, we can write
\begin{equation}
    \ket{S_{0,\,0}} = e^{\frac{\beta_0}{2}\sum_{i\in \Lambda_0} X_i}e^{\frac{\beta_1}{2}\sum_{i\in \Lambda_1} X_i}\, \ket{\text{Cluster}}\,,
\end{equation}
with $\ket{\text{Cluster}} = \prod_i CZ_{i,\,i+1}\ket{+\dots+}$ is the standard SPT cluster state, $CZ_{j,\,k}$ is a controlled-Z unitary gate and $\ket{+\dots+}$ is the standard product state in the $X$ basis, with positive eigenvalues. 

To establish $\ket{S_{0,\,0}}$ is indeed a area-law state, it is convenient to represent it in its matrix product state (MPS) form, which can be built out of the matrices
\begin{align}
    A^0_e &= \begin{pmatrix}
        \sinh{\frac{\beta_0}{2}}&\sinh{\frac{\beta_0}{2}}\\[0.3em]
        \cosh{\frac{\beta_0}{2}}&\cosh{\frac{\beta_0}{2}}
    \end{pmatrix}\,, \nonumber\\
    A^1_e &= \begin{pmatrix}
        \cosh{\frac{\beta_0}{2}}&-\cosh{\frac{\beta_0}{2}}\\[0.3em]
        \sinh{\frac{\beta_0}{2}}&-\sinh{\frac{\beta_0}{2}}
    \end{pmatrix}\,,
\end{align}
and similarly for $\Lambda_o$ with the exchange $\beta_0\Leftrightarrow \beta_1$. The existence of an exact MPS representation automatically ensures that these states satisfy an entanglement area-law instead of volume law \cite{Cirac2021, Verstraete2023, Tantivasadakarn2024}. 

To insert a symmetry defect, we use the following setup: Given a local Hamiltonian $H=\sum_j h_j$ invariant under a global symmetry with onsite action $G=\prod_{i\in \Lambda} g_i$, we can define a new Hamiltonian $\Tilde{H} = \sum_{j\neq k}h_j+\Tilde{h}_k$, where $\Tilde{h_k} = g_k\,h_k\,g^{-1}_k$. This modification corresponds to inserting a localized symmetry defect, i.e., the defect Hamiltonian differs from the original one solely by the conjugation of a single term by the local symmetry operator \cite{Grimm2003}. For the $\mathbb{Z}_2$ symmetries we will be interested, this amounts to flipping the sign of terms across the defect boundary. The spatial location of the symmetry defect is irrelevant, since it can be moved by means of global unitary operators \cite{Seiberg2024}.

Following this prescription, we consider the modified Hamiltonian
\begin{align}
    \tilde{H}^{(2)}_{s} = \sum_{a}\sum_{j\neq K}\alpha_{a,\,j}\left(e^{-\beta_a\left(X_j+X_{j+2}\right)} - Z_jX_{j+1}Z_{j+2}\right)\nonumber\\
    +\alpha_{a,\,K}\left(e^{-\beta_{a}\left(X_K+X_{K+2}\right) } + Z_KX_{K+1}Z_{K+2}\right)\,,
\end{align}
where the defect is placed at site $K$ (which can be an even or odd site). Generally, inserting a symmetry defect is usually enough to destroy quantum many body scarring \cite{Fontana2025}, since scarring typically depends on a delicate cancellation between the Hamiltonian terms. In this model, however, we can show that scarring survives. To make this explicit, we apply the unitary transformation $V_1 \tilde{H}^{(2)}_{s} V_1^\dagger$, with $V_1=\exp(\frac{i\pi}{4}\sum_j Z_jZ_{j+1})=\prod_j CZ_{j,\,j+1}$, yielding 
\begin{widetext}
\begin{align}
V_1\tilde{H}^{(2)}_{s}V_1^\dagger = \sum_{a}\sum_{j\neq K}\alpha_{a,\,j}&\left(e^{-\beta_a\left(Z_{j-1}X_jZ_{j+1}+Z_{j+1}X_{j+2}Z_{j+3}\right)} - X_{j+1}\right)\nonumber\\
    &+\alpha_{a,\,K}\left(e^{-\beta_{a}\left(Z_{K-1}X_KZ_{K+1}+Z_{K+1}X_{K+2}Z_{K+3}\right) } + X_{K+1}\right)\,,
    \label{eq: transformed def hamiltonian}
\end{align}
\end{widetext}

from \eqref{eq: transformed def hamiltonian}, one can readily verify that the state
\begin{equation}
    \ket{\tilde{S}_{(t_K)}} = \left[\prod_a e^{\frac{\beta_a}{2}\sum_{i\in \Lambda_a}Z_{j-1}X_jZ_{j+1}}\right]\ket{-}_{K+1}\ket{X_+}\,,
\end{equation}
with $\ket{X_+}=\otimes_{j\neq K+1}\ket{+}_j$. This state is annihilated by $V_1\tilde{H}^{(2)}_sV_1^\dagger$. Undoing the unitary transformation yields the desired state, which is annihilated by the original $\tilde{H}^{(2)}_S$:
\begin{equation}
V_1\ket{\tilde{S}_{(t_K)}} = \left[\prod_a e^{\frac{\beta_a}{2}\sum_j X_j}\right]V_1 \ket{-}_{K+1}\ket{X_+}\,,    
\end{equation}
the state $V_1\ket{\tilde{S}_{(t_K)}}$ is a zero-energy state of the Hamiltonian $\tilde{H}^{(2)}_s$, and retains an area-law entanglement entropy, as it can still be represented as a low bond-dimension MPS state. Similarly, we can construct the corresponding scar state in the two defect sector. By induction, we obtain
\begin{align}
    \ket{\tilde{S}_{(t_{K-1} , t_{K})}} &= \left[\prod_a e^{\frac{\beta_a}{2}\sum_j X_j}\right]V_1\ket{-}_{2K+1}\ket{-}_{2K+2}\ket{X^{\prime}_+}\,,\\ \nonumber
    \quad \ket{X^{\prime}_+}&=\underset{i\neq K,\,K+1}{\otimes}\ket{+}_i. 
\end{align}
\begin{figure}
    \centering
    \subfloat[\label{fig:2-chain 1 defect sector}]{\includegraphics[width=\linewidth]{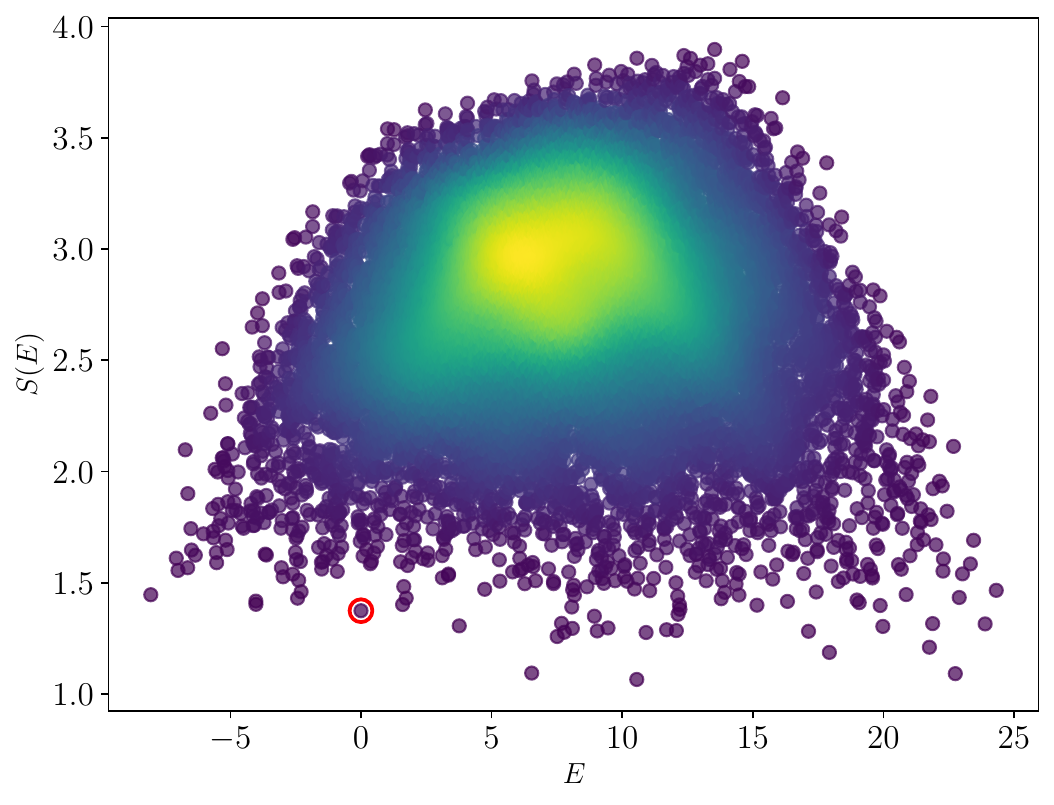}}
    \vspace{1pt}
    \subfloat[\label{fig:2-chain 2defect sector}]{\includegraphics[width=\linewidth]{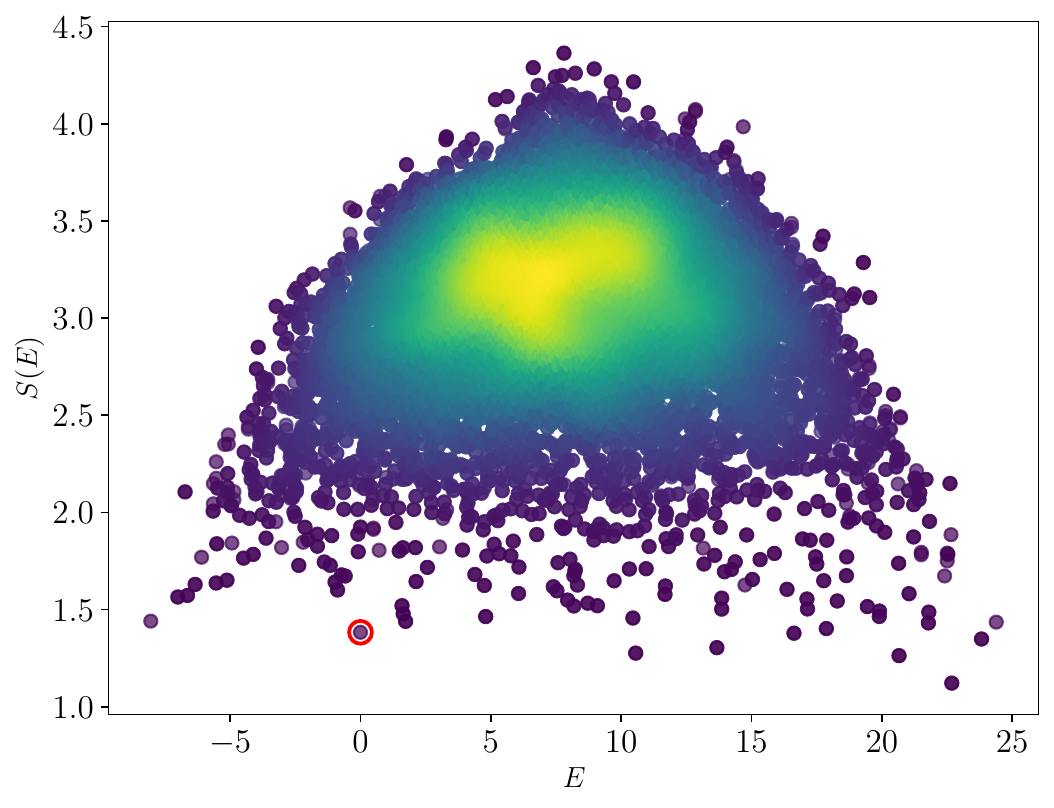}}
    \caption{\justifying Entanglement entropy profile in the corresponding sectors with $\mathbb{Z}_2$ defect insertion. (a) The profile corresponding to the sector with a single defect at site $j=7$ and (b) the sector with the insertion of two defects (sites, $j=7, 10$) on each of the corresponding sub-lattices. In both situations, the scarring is preserved and we are able to identify the single QMBS state at zero-energy, as depicted in the figure by the red circle. These were obtained from the exact diagonalization of the corresponding model over a periodic chain of size $L=14$ sites. The other parameters of the model were fixed as $\beta=0.5$ and $\alpha=0.3$.}
    \label{fig:placeholder}
\end{figure}
The existence of the exact scar eigenstates can be verified numerically, as we show in Fig.\ref{fig:2-chain 1 defect sector} and Fig.\ref{fig:2-chain 2defect sector} where we observe that these are all low-entangled states at zero energy. Although our discussion focused on the $k=2$ chain, this behavior is shared by a broader class of these chains. Specifically, under duality transformations, every QMBS state is mapped to a corresponding dual QMBS state residing in another twisted sector of the Hilbert space. This remarkable property, allows these chains to preserve non-thermal features (in this case, scarring) even under complicated transformations, non-invertible transformation such as the Kramers-Wannier and Kennedy-Tasaki transformations. In fact, the compatibility of QMBS embedding holds for all $k$-chain with $k\in 2\mathbb{Z}$, as we show in the upcoming section Sec.\ref{subsectionC}.

\subsection{the $k=3$-chain}

We now turn our attention to the $3$-chain. The model is defined by
\begin{equation}
    H^{(3)}_s=\sum_{j,\,a}\alpha^{(3)}_{j,\,a}\left(e^{-\beta_a\left(X_{j}+X_{j+3}\right)}-Z_{j}X_{j+1}X_{j+2}Z_{j+3}\right)\,.
\end{equation}
The coupling is defined by
$$
\alpha^{(3)}_{j,a}
=
|\alpha|
+
(-1)^{\frac{j-a}{3}},
\qquad
j=a\pmod 3.
$$
Thus, within each sublattice, the coupling alternates between
$$
\alpha_+ = |\alpha|+1,
\qquad
\alpha_- = |\alpha|-1,
$$
with $|\alpha|<1$. As before, $a=0,\,1,\,2$ is a sublattice index. Unlike the $k=2$ chain, the current model display a degenerate subspace. This degeneracy arises from the constraint $\prod_j C^{(3)}_j=\openone$, which implies that not all states stabilized by $C^{(3)}$ are not mutually independent.
This degenerate scar subspace is connected by the global $\mathbb{Z}_2$ symmetry of the model and can be written as
\begin{equation}
    \ket{S^{(3)}_a} = \left[\prod_a e^{\frac{\beta_a}{2}\sum_j X_j}\right]\ket{C^{(3)}}\,,
\end{equation}
where $\ket{C^{(3)}}$ is the state stabilized by the operator $Z_{j}X_{j+1}X_{j+2}Z_{j+3}$, meaning it has a positive unity eigenvalue. The $\ket{C^{(3)}}$ state, can be expressed in a simple form by considering a pivot transformation with an SPT phase \cite{Tantivasadakarn2023} via the unitary operator
\begin{equation}
    V_2 = e^{\frac{i\pi}{4}\sum_j Z_jX_{j+1}Z_{j+2}}\,.
\end{equation}

\begin{figure}
    \centering
    \subfloat[\label{fig: 3-chain no defect}]{\includegraphics[width=\linewidth]{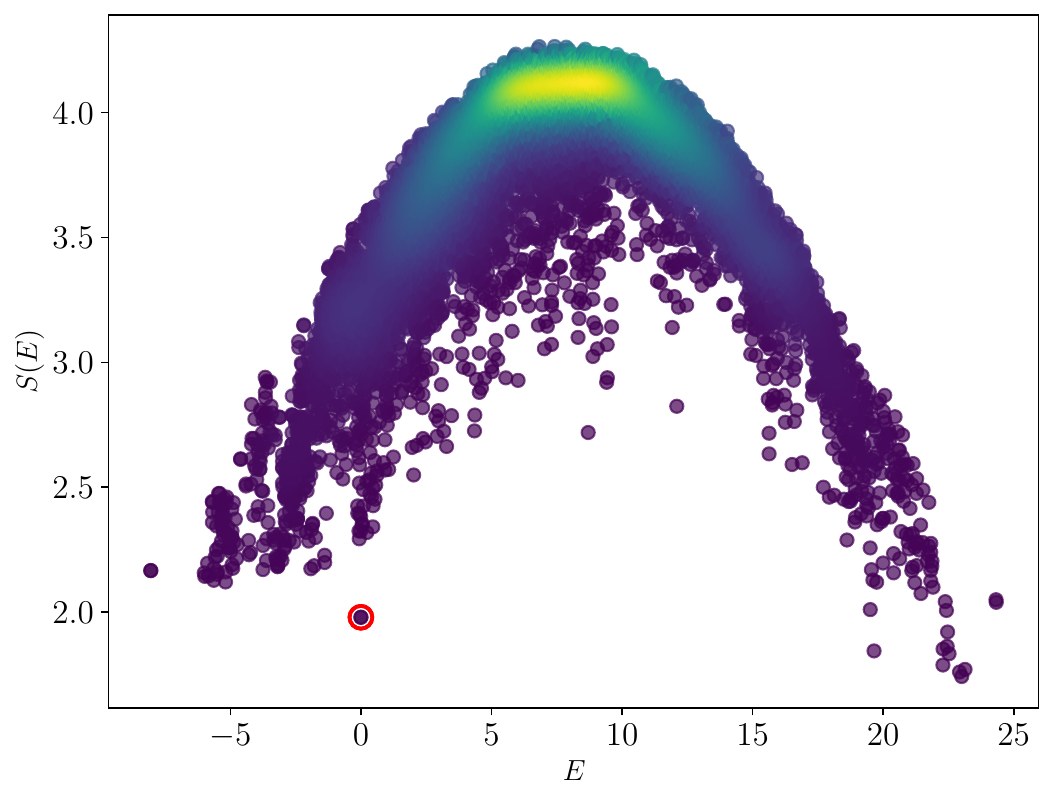}}
    \vspace{1pt}
    \subfloat[\label{fig: 3-chain 1 defect}]{\includegraphics[width=\linewidth]{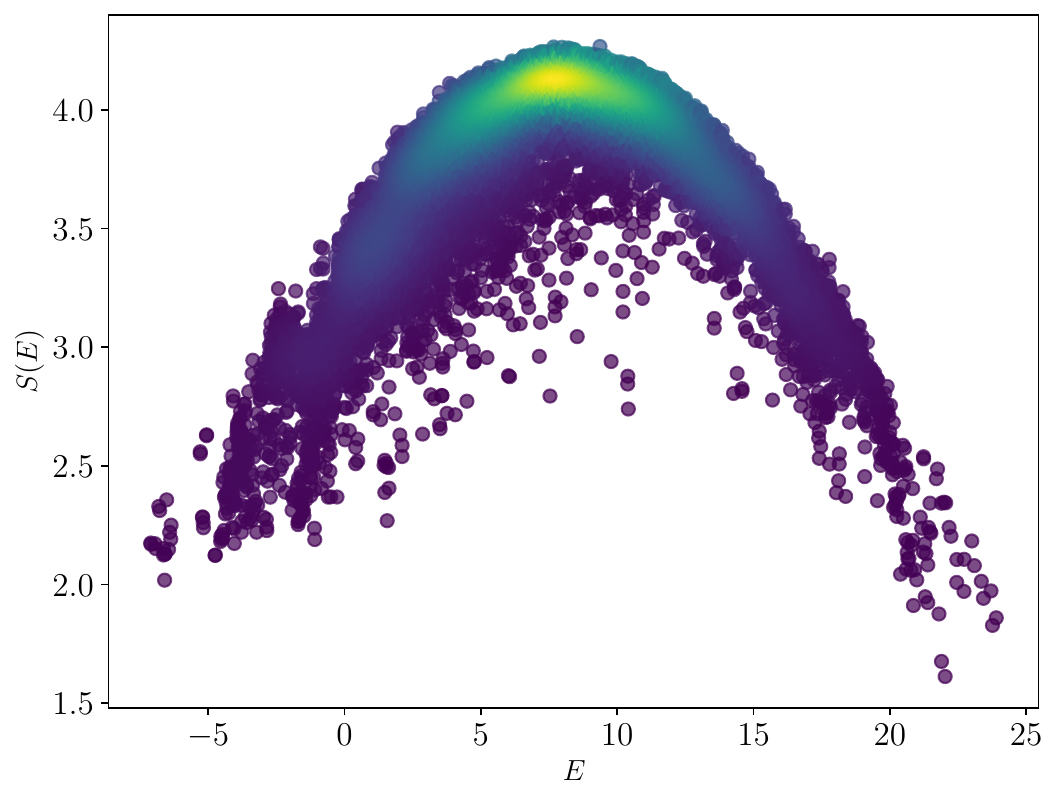}}
    \caption{\justifying Entanglement entropy profile for the scarred $3$-chain. (a) The profile corresponding to the sector with zero defects. (b) The profile corresponding to the sector with a single defect at $j=7$. Unlike the previous case, inserting a defect for this chain, completely destroy the scarring of the model, as predicted by our analytical analysis. These were obtained from exact diagonalization of the corresponding model for a periodic chain of size $L=14$ sites. The other parameters of the model were fixed as $\beta=0.5$ and $\alpha=0.3$.}
    \label{fig:placeholder}
\end{figure}

Following a strategy similar to the previous section, we define the transformed Hamiltonian $\mathfrak{h}^{(3)}_s = V_2H^{(3)}_sV_2^\dagger$. The effect of such transformation, is the $C^{(3)}\mapsto C^{(1)}$ and $C^{(0)}\mapsto C^{(4)}$. Therefore, the $\ket{C^{(3)}}$ state can be explicitly represented by the action of the pivot $V_2$ on top of the state stabilized by the set of operators  $\{C^{(1)}_i\} = \{Z_iZ_{i+1}\}$, i.e., 
\begin{equation}
    \ket{C^{(3)}_{\pm}} = e^{\frac{i\pi}{4}\sum_j Z_{j}X_{j+1}Z_{j+2}}\left(\otimes_j \ket{0}_j\pm \otimes_j \ket{1}_j\right)\,,
\end{equation}
where $\ket{0},\,\ket{1}$ denote the computational states in the $Z$-basis. Unlike the $k=2$ case, scarring in the present situation only survives in the untwisted sector (i.e., in the absence of defects). The insertion of a $\mathbb{Z}_2$ defect spoils the stabilizer condition $C^{(k)}\ket{C^{(k)}}=\ket{C^{(k)}}$ necessary for the embedding. Indeed, inserting a $\mathbb{Z}_2$ defect, would be equivalent (after pivoting with $V_2$) to flipping the sign of a single Ising term, This sign flip violates the global stabilizer condition, meaning that no state exists that can simultaneously stabilize the updated stabilizer set $\{-Z_KZ_{K+1}\}\cup\{ Z_iZ_{i+1}\}_{i\neq K}$. Consequently, defect insertion completely destroys the quantum many body scarring. This can also be verified numerically from the entanglement entropy profile of the model, as depicted in Fig.\ref{fig: 3-chain no defect} and Fig.\ref{fig: 3-chain 1 defect}.

\subsection{General case \label{subsectionC}}
Having analyzed these explicit examples, we are now in position to discuss the general $k$-chain classification. It is in fact, a very simple observation. For every instance of $k\in \mathbb{Z}_{\text{even}}$, the corresponding $\alpha$-chain will have robust many body scarring phenomena because the anchor state (the stabilizer state) stems from a SPT ground state. Conversely, for $k\in \mathbb{Z}_{\text{odd}}$, scarring will be destroyed upon insertion of defects, due to the non-trivial nature of the anchor state will violate the stabilizer conditions required to embed scars.

For a generic $k$-chain, the pivoting framework \cite{Tantivasadakarn2023} yields the general transformation
\begin{equation}
    U_{\Bar{\alpha}} H^{(k)} U^\dagger_{\Bar{\alpha}} = H^{(2\Bar{\alpha} -k)}\,, ~ ~ ~ U_{\Bar{\alpha}} = e^{\frac{i\pi}{4}H^{(\Bar{\alpha})}}\,.
\end{equation}

The models preserving the embedding conditions under defect insertions, are those that can be mapped to a trivial model (i.e., $H^{(0)}$) under a rotation around the pivot $H^{\bar{\alpha}}$. This imposes the constraint that the required pivot is the one with $\Bar{\alpha} = k/2$, with $k,\,\Bar{\alpha}\in \mathbb{Z}$. This immediately implies that the models preserving the scar embedding in the presence of defects are those with $k \in \mathbb{Z}_{\text{even}}$, as odd chains ($k\in \mathbb{Z}_{\text{odd}}$) do not yield the pivot index $\Bar{\alpha}$ an integer.

For those models with $k\in \mathbb{Z}_{odd}$, they can always be connected to the classical Ising model, described by the $k=1$, which can then be trivialized via a Kramers-Wannier duality. This implies that for $k\in\mathbb{Z}_{odd}$, the embedding is preserved only within the symmetric sector of the scar states. We highlight that the preservation of scarring properties perfectly aligns with the state classification of the underlying anchor states: namely, whether the embedded states originates from an SPT phase $(k\in \mathbb{Z}_{even})$ or a symmetry broken phase $(k \in \mathbb{Z}_{odd})$ \cite{Verresen2017}. 

\section{Scar preserving deformation in the presence of a duality Interface \label{secIV}}

\subsection{detailed half-gauging of the scarred 1-chain}
Up to now, we have focused on wether or not scarring could survive in the presence of symmetry defects. We have shown the conditions under which scarred states can persist within our setup. Another interesting direction to explore regards the fate of the QMBS states upon insertion of a \textit{duality defect} (more precisely for our case, a \textit{duality interface}, since the duality will not be a symmetry of the system in the current analysis.). This object is obtained by truncating the duality operation at a specific real space site, say a site labeled by $J$. This procedure separates the system in two parts: one governed by the original Hamiltonian and the other by its dual, with both regions ``glued'' together by an appropriate interface term. We will show that scarring can still survive in these scenarios. Furthermore, we find that the scarred subspace now acquires a degeneracy that is protected by new symmetries induced due to the presence of the interface. 

The strategy in this section is to explore the $1$-chain via half-gauging \cite{Choi2022, Choi2023, Sinha2023, Seiberg2024, Yamazaki2024}. Specifically, we gauge the global $\mathbb{Z}_2$ symmetry up to some arbitrary site $J$ of the chain. In doing so, we obtain a scenario in which half of the system is described by the original model, and the other half is described by its dual, with both descriptions separated by a so-called \textit{duality interface}. The precise position of the interface is immaterial, as it can be moved by using an appropriate unitary operation. While we focus on the scarred 1-chain as a concrete example, this construction applies to the entire family of $k$-chains, because it only relies on gauging the corresponding symmetry. In fact, it is not even restricted to one dimension. 

To build intuition, we first describe the procedure on a periodic chain. Consider the $1$-chain defined over a periodic one-dimensional lattice, with sites ranging from $1$ to $L$. Periodicity is enforced via the identification
\begin{equation}
    Z_{L+1}=Z_1\,, ~~~ X_{L+1} = X_1\,.
\end{equation}

Next, we introduce gauge variables on the links of the lattice, labeled by half-integer sites, up to some arbitrary site $J$, we introduce new spin variables $\tau^x_{i-\frac{1}{2}}$ and $\tau^z_{i-\frac{1}{2}}$. This setup yields the following gauged model:
\begin{align}
    H^{(1)}_{sG} &= \alpha_L\left(e^{-\beta}\left(X_L+X_1\right) - Z_L\tau^x_{\frac{1}{2}}Z_1\right)\nonumber\\
    &+\sum_{i=1}^{J-2}\alpha_1\left(e^{-\beta\left(X_i+X_{i+1}\right)} - Z_i\tau^x_{i+\frac{1}{2}}Z_{i+1}\right)\nonumber\\
    &+\alpha_{J-1}\left(e^{-\beta\left(X_{J-1}+X_J\right)} - Z_{J-1}\tau^x_{J-\frac{1}{2}}Z_J\right)\nonumber\\
    &+\sum_{i=J+1}^{L-1}\alpha_i\left(e^{-\beta\left(X_i+X_{i+1}\right)} -Z_iZ_{i+1}\right)\,.
\end{align}

Clearly, this model possesses local conserved quantities (Gauss laws) given by
\begin{equation}
    G_i = \tau^z_{i-\frac{1}{2}}X_i\tau^z_{i+\frac{1}{2}}\,, ~ ~ ~ i=1,\,\dots,\,J-1\,,
\end{equation}
due to the periodic boundary conditions, the global $\mathbb{Z}_2$ symmetry is recovered from the products of the local Gauss laws, i.e., $\prod_i G_i = \prod_i X_i$. Notice that inserting these gauge variables expands the Hilbert space by a factor of $2^J$, yet we only obtain $J-1$ Gauss law constraints. This mismatch hints that the interface must exhibit some non-trivial behavior in order to accommodate the extra degree of freedom.  

We now project the gauged model onto the physical subspace by enforcing the Gauss law constraint $G_i=+1$. This yields the operator identity $X_i=\tau^z_{i-\frac{1}{2}}\tau^z_{i+\frac{1}{2}}$, allowing us to express the Hamiltonian purely in terms of the gauge fields
\begin{align}
    H^{(1)}_{sG} &= \alpha_L\left(e^{-\beta}\left(X_L+\tau^z_{\frac{1}{2}}\tau^z_{\frac{3}{2}}\right) - Z_L\tau^x_{\frac{1}{2}}Z_1\right)\nonumber\\
    &+\sum_{i=1}^{J-2}\alpha_1\left(e^{-\beta\left(\tau^z_{i-\frac{1}{2}}\tau^z_{i+\frac{1}{2}}+\tau^z_{i+\frac{1}{2}}\tau^z_{i+\frac{3}{2}}\right)} - Z_i\tau^x_{i+\frac{1}{2}}Z_{i+1}\right)\nonumber\\
    &+\alpha_{J-1}\left(e^{-\beta\left(\tau^z_{J-\frac{3}{2}}\tau^z_{J-\frac{1}{2}}+X_J\right)} - Z_{J-1}\tau^x_{J-\frac{1}{2}}Z_J\right)\nonumber\\
    &+\sum_{i=J+1}^{L-1}\alpha_i\left(e^{-\beta\left(X_i+X_{i+1}\right)} -Z_iZ_{i+1}\right)\,,
\end{align}
To bring the Hamiltonian into a more appealing form, it is convenient to introduce the following gauge invariant variables
\begin{alignat}{2}
    \hat{Z}_{i+\frac{1}{2}} &= \tau^z_{i+\frac{1}{2}}\,,~ ~ ~ &i&=1\,,\dots\,,J-1\,,\\
    \hat{X}_{i+\frac{1}{2}} &= Z_i\tau^x_{i+\frac{1}{2}}Z_{i+1}\,, ~ ~ ~ &i&=1,\,\dots,\,J-1\,,\\
    \hat{X_{\frac{1}{2}}} &= \tau^x_{\frac{1}{2}}Z_1\,,~ ~ ~ &\hat{X}_{J-\frac{1}{2}} &= Z_{J-1}\tau^x_{J-\frac{1}{2}}\,,
\end{alignat}
in terms of these hatted variables, the Hamiltonian takes the more compact form
\begin{align}
       H^{(1)}_{sG} &= \alpha_L\left(e^{-\beta\left(X_L + \hat{Z}_{\frac{1}{2}}\hat{Z}_{\frac{3}{2}}\right)} - Z_L\hat{X}_{\frac{1}{2}}\right)\nonumber\\
    &+\sum_{i=1}^{J-2}\alpha_i\left(e^{-\beta\left(\hat{Z}_{i-\frac{1}{2}}\hat{Z}_{i+\frac{1}{2}}+\hat{Z}_{i+\frac{1}{2}}\hat{Z}_{i+\frac{3}{2}}\right)} - \hat{X}_{i+\frac{1}{2}}\right)\nonumber\\
    &+\alpha_{J-1}\left(e^{-\beta\left(\hat{Z}_{J-\frac{3}{2}}\hat{Z}_{J-\frac{1}{2}} + X_J\right)} - \hat{X}_{J-\frac{1}{2}}Z_J\right)\nonumber\\
    &+\sum_{i=J}^{L-1}\alpha_i\left(e^{-\beta\left(X_{i}+X_{i+1}\right)} -Z_{i}Z_{i+1}\right)\,.
    \label{eq: mid step hamilt 31}
\end{align}
 
Finally, we can rewrite the above Hamiltonian using a single set of Pauli operators by means of a new set of variables
\begin{alignat}{2}
    \sigma^z_{i+1} &= \hat{Z}_{i+\frac{1}{2}}\,, ~ ~ ~ &i&=1,\,\dots,\,L\,,\\
    \sigma^x_{i+1} &= \hat{X}_{i+\frac{1}{2}}\,, ~ ~ ~ &i&=1,\,\dots,\,L\,,\\
    \sigma^z_{1} &= \hat{Z}_{\frac{1}{2}}\,, ~ ~  &\sigma^x_{1} &= \hat{X}_{\frac{1}{2}}\,,\\
    \sigma^z_{\ell} &= \hat{Z}_{J-\frac{1}{2}}\,, ~ ~ ~ &\sigma^x_{\ell} &= \hat{X}_{J-\frac{1}{2}}\,.
\end{alignat}

It is important to stress a detail about the above redefinition of variables. To preserve the correct algebraic relations among the bonds in \eqref{eq: mid step hamilt 31}, we unavoidably need to consider a modified lattice which contains one extra site between the original $J-1$ and $J$ sites. This requirement stems from our earlier observation that, even after projection onto the physical Hilbert space, the gauged Hilbert space cannot be reduced back to its original dimension. In this construction, the remaining degree of freedom manifests as a \textit{translation defect} (following the nomenclature proposed in \cite{Seiberg2024}). Had we not introduced this extra site, we would be unable to write the Hamiltonian \eqref{eq: mid step hamilt 31} using solely a single set of spin-$\frac{1}{2}$ operators while preserving the appropriate bond algebra \cite{Cobanera2011}. 

The final Hamiltonian simplies to

\begin{align}
    H^{(1)}_{sG} &= \alpha_L\left(e^{-\beta(\sigma^x_L+\sigma^z_1\sigma^z_2)} - \sigma^z_L\sigma^x_1\right)\nonumber\\ 
    &+\sum_{i=1}^{J-2}\alpha_i \left(e^{-\beta\left(\sigma^z_{i}\sigma^z_{i+1}+\sigma^z_{i+1}\sigma^z_{i+2}\right)}-\sigma^x_{i+1}\right)\nonumber\\
    &+\alpha_{J-1}\left(e^{-\beta\left(\sigma^z_{J-1}\sigma^z_{\ell}+\sigma^x_J\right)} - \sigma^x_\ell\sigma^z_{J} \right)\nonumber\\
    &+\sum_{i=J}^{L-1}\alpha_j\left(e^{-\beta\left(\sigma^x_i+\sigma^x_{i+1}\right)}-\sigma^z_{i}\sigma^z_{i+1}\right)\,. 
    \label{eq: mixed scars}
\end{align}

This is the desired result. The half-gauged Hamiltonian splits the system into two regions, each part described by its respective model (original or dual), separated by the interface term $\sigma^x_\ell \sigma^z_J$. This analysis demonstrates that to study duality defects in the $1$-chain, it suffices to modify the boundary term between the two phases in this manner. Because the interface is topological, it can be moved by means of an appropriate unitary operation. Consequently, pairs of interfaces can be fused \cite{Oshikawa2023,Seiberg2024, Shao2024}, as a result, depending on the fusion channel, this yields another Hamiltonian that contains: (i) no defects or (ii) a $\mathbb{Z}_2$ symmetry defect, both of which have already been explored in previous sections. For that reason, we are more interested in the situation in which we have only a single interface to determine whether non-trivial scarring can persist. In what follows, we leverage this analysis to define a new Hamiltonian containing only a single interface term
\begin{align}
    \mathfrak{H}^{(1)}_{sG} &= \sum_{i=2}^{J-2}\alpha_{i}\left(e^{-\beta\left(\sigma^z_{i-1}\sigma^z_i+\sigma^z_i\sigma^z_{i+1}\right)}-\sigma^x_i\right)\nonumber\\
    &+\alpha_{J-1}\left(e^{-\beta\left(\sigma^z_{J-2}\sigma^z_{J-1}+\sigma^x_J\right)}-\sigma^x_{J-1}\sigma^z_J\right)\nonumber\\
    &+\sum_{i=J}^{L-1}\alpha_i\left(e^{-\beta\left(\sigma^x_i+\sigma^x_{i+1}\right)}-\sigma^z_{i}\sigma^z_{i+1}\right)\,.
    \label{eq: 1duality defect hamiltonian}
\end{align}

Here, a single interface at a site $J$ separates the two distinct descriptions of the model. Since a periodic chain cannot accommodate a single duality interface, in \eqref{eq: 1duality defect hamiltonian} we impose open boundary conditions. The symmetry content of the Hamiltonian \eqref{eq: 1duality defect hamiltonian} is particularly interesting. First, notice that there are two non-local operators $\eta_1$ and $\eta_2$, defined as
\begin{equation}
    \eta_1 = \left(\prod_{j=1}^{J-1}\sigma^x_j\right)\,, ~~~\eta_2 = \sigma^z_{J-1}\prod_{j=J}^LX_j\,,
    \label{eq: categorical symmetries}
\end{equation}
 these operators commute with the Hamiltonian, $\left[\mathfrak{H}^{(1)}_{sG},\, \eta_{1,\,2}\right]=0$, but anticommute with each other
\begin{equation}
    \eta_1\eta_2 = -\eta_2\eta_1\,,
    \label{eq: anomaly}
\end{equation}
this algebraic structure is characteristic of an anomaly between the symmetry operators and signals non-trivial physics. Specifically, because the irreducible representation of the $\eta_1$ and $\eta_2$ algebra is at least two-dimensional, the entire spectrum of $\mathfrak{H}^{(1)}_{sG}$ exhibits a two-fold symmetry protected degeneracy. Consequently, the QMBS subspace becomes four-fold degenerate (rather than two-fold, as in the interface free case) to remain compatible with the anticommutation of $\eta_1$ and $\eta_2$.

We emphasize that, while the $1$-chain serves as an illustrative example, this behavior is not a particularity of this model; rather, it is a direct consequence of the presence of the interface itself. This can be argued from the fact that, since the interface separates both the original and dual descriptions, the symmetries on both sides will also follow the corresponding description. In the Ising model language, we can think of the symmetries \eqref{eq: categorical symmetries} as the \textit{categorical symmetries} of the mixed representation of the underlying model \cite{Ji2020}. Thus, naturally, if we are to preserve the bond-algebra of the model, $\eta_1$ and $\eta_2$ have to anticommute. In fact, considering the Kramers-Wannier mapping
\begin{align}
X_i\openone_{i+1}&\mapsto \tau^z_{i-\frac{1}{2}}\tau^z_{i+\frac{1}{2}}\,,\\
Z_iZ_{i+1}&\mapsto \openone_{i-\frac{1}{2}}\tau^x_{i+\frac{1}{2}}\,,
\end{align}
it becomes clear that the anticommutation between the original bonds $\left\{X_i\openone_{i+1},\,Z_iZ_{i+1}\right\}=0$ induces the anticommutation between the mixed ones $\left\{X_i\openone_{i+1},\, \openone_{i-\frac{1}{2}}\tau^x_{i+\frac{1}{2}}\right\}=0$. Therefore, the anomaly observed in \eqref{eq: anomaly} stems from the mutual statistics between the original and dual descriptions, and therefore, it should be present in any half-gauged model. In fact, in the appendices, we show that other chains (like the 2-chain and 3-chain) also exhibit anomalous algebras upon insertion of the duality interface. This unavoidably increases the degeneracy of the scarred subspace and also provides a symmetry protection mechanism to it. The same conclusion holds in higher dimensions. 
\begin{figure}[t]
    \centering
    \begin{tikzpicture}

        \node[inner sep=0pt] (main) {
            \includegraphics[width=\columnwidth]%
            {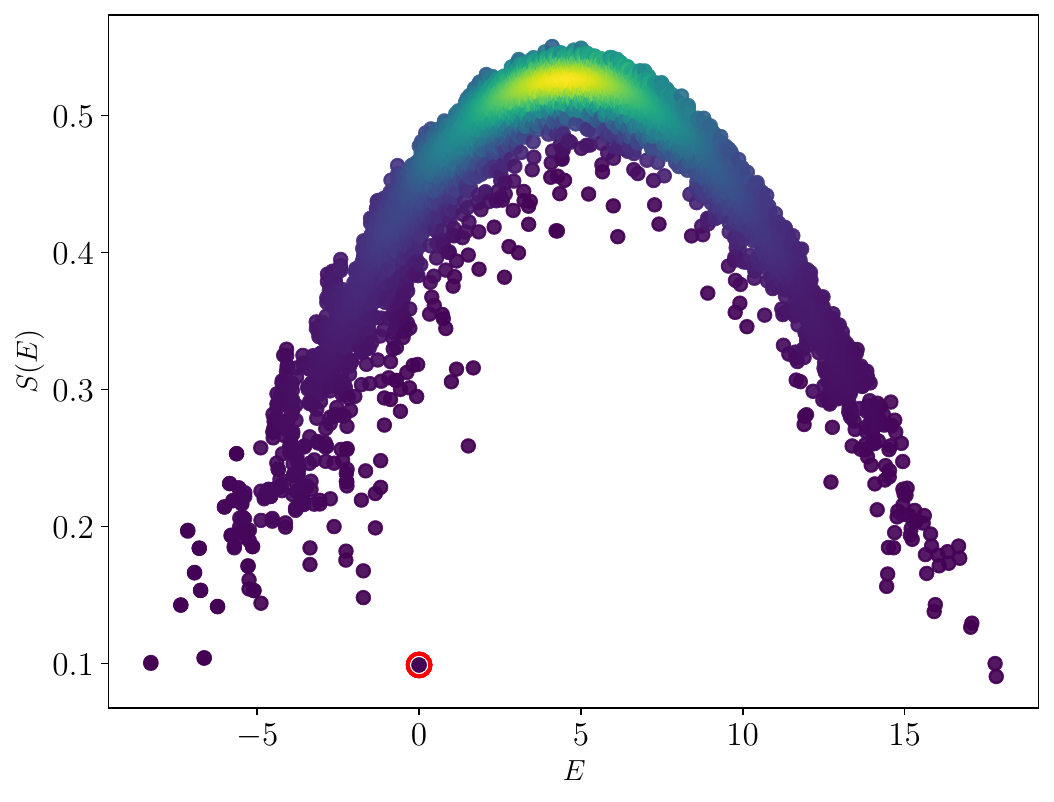}
        };

        \node[
            draw=black,
            line width=0.5pt,
            inner sep=0pt,
            anchor=south west
        ] at ([xshift=0.35\columnwidth,yshift=0.20\columnwidth]main.south west) {
            \includegraphics[width=0.38\columnwidth]%
            {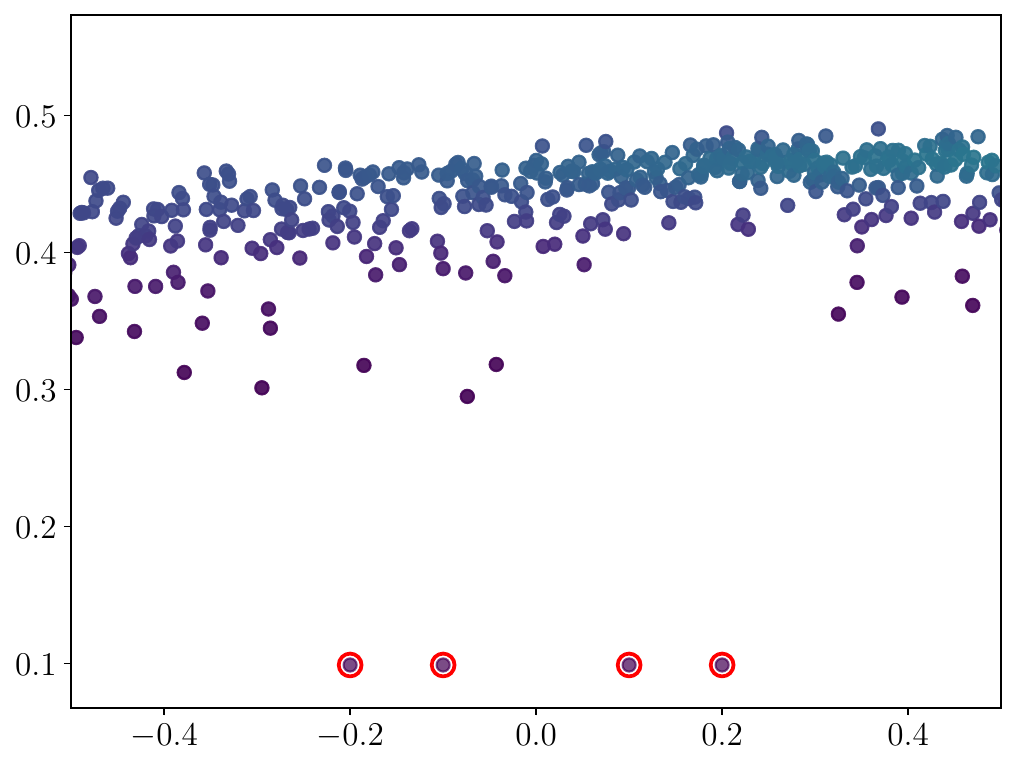}
        };

    \end{tikzpicture}

    \caption{\justifying
    Entanglement profile of the 1-chain containing a single duality interface.
    Due to the anomaly, the QMBS subspace increases from a 2-fold degeneracy to
    a 4-fold degeneracy, which is now protected by the underlying symmetries.
    In the inset, we split the states in energy by hand, simply to make
    explicitly the presence of the four states. These were obtained by means of
    exact diagonalization of an open chain of $L=14$ sites. The parameters were
    fixed as $\beta=0.5$ and $\alpha=0.3$.
    }
    \label{fig:single_duality_defect_SvsE}
\end{figure}

For completeness, the exact scarred states for the Hamiltonian \eqref{eq: 1duality defect hamiltonian}, take the form
\begin{equation}
    \ket{\mc S_{\pm}}=e^{\frac{\beta}{2}\sum_{j=1}^{J-1}\sigma^z_j\sigma^z_{j+1}}e^{\frac{\beta}{2}\sum_{j=J}^{L-1}\sigma^x_j}\ket{\psi_{\pm}}\,, \label{eq: scarred state mixed chain}
\end{equation}
where $\ket{\psi_{\pm}}$ are defined as
\begin{align}
    \ket{\psi_{+}}=\ket{\varsigma^x_1}\ket{+_2\dots+_{J-2}}\ket{+_{J-1}0_J}\ket{0_{J+1}\dots 0_{L}}\,,\\
    \ket{\psi_{-}} = \ket{\varsigma^x_{1}}\ket{+_2\dots +_{J-2}}\ket{-_{J-1}1_J}\ket{1_{J+1}\dots1_{L}}\,,
\end{align}
where $\ket{\varsigma_x}$ is a $\sigma^x$ eigenstate, and $\ket{0}$ and $\ket{1}$ are the up/down eigenstates in the $\sigma^z$-basis.

\subsection{Half-gauging of the 2-chain with OBC}

We briefly discuss half-gauging of the 2-chain to illustrate the loss and
restoration of a protected-mode interpretation of the QMBS degeneracy.
Details of the construction are provided in
Appendix~\ref{app: hg 2-chain}.

The model is defined as
\begin{equation}
    H^{(2)}_{s}=\sum_{j=1}^{L-2}\alpha_j\left(e^{-\beta\left(X_j+X_{j+2}\right)}-Z_jX_{j+1}Z_{j+2}\right)\,.
    \label{eq:44}
\end{equation}
For $\beta=0$, Eq.~\eqref{eq:44} reduces, up to an additive constant,
to the cluster model with OBC, a stabilizer model with symmetry-protected
edge modes \cite{Kitaev2011, Cirac2011, Xie2011, Nayak2014,Sarma2015, Alicea2016}.
The corresponding edge operators are
\begin{alignat}{2}
    \chi_1&=X_1Z_2\,, ~ ~ ~ &\chi_2& = Z_1\,,\\
    \zeta_1&=Z_{L-1}X_L\,, ~ ~ ~ &\zeta_2& = Z_L\,,
\end{alignat}
which commute with the Hamiltonian $H^{(2)}_s(0)$ and anticommute
within each edge pair, i.e.,
$\{\chi_1,\,\chi_2\}=\{\zeta_1,\,\zeta_2\}=0$.
These operators define one qubit at each edge and account for the
fourfold degeneracy of the model. When $\beta\neq0$, a four-dimensional
subspace remains embedded in the middle of the spectrum, owing to the
absence of boundary cluster stabilizers, but the original edge operators
are no longer conserved. The scarred subspace therefore remains fourfold
degenerate but need not retain the same protection relevant to quantum
computation \cite{Sarma2015, Alicea2016}. We now half-gauge the model to
recover a protected qubit in the scarred system. Its conserved operators
are no longer confined to an edge but instead extend from the interface
introduced by half-gauging.

The half-gauged model can be written as
\begin{align}
H^{(2)}_{sG}&=\alpha_1\left(e^{-\beta\left(\sigma^z_2+\sigma^z_{2}\sigma^z_4\right)}-\sigma^z_1\sigma^x_2\sigma^z_3\right)\nonumber\\
&+\sum_{j=2}^{J-3}\alpha_j\left(e^{-\beta\left(\sigma^z_{j-1}\sigma^z_{j+1}+\sigma^z_{j+1}\sigma^z_{j+3}\right)}-\sigma^z_{j}\sigma^x_{j+1}\sigma^z_{j+2}\right)\nonumber\\
&+\alpha_{J-2}\left(e^{-\beta\left(\sigma^z_{J-3}\sigma^z_{J-1}+\sigma^z_{J-1}\sigma^x_{J}\right)}-\sigma^z_{J-2}\sigma^x_{J-1}\sigma^z_{\ell}\right)\nonumber\\
&+\alpha_{J-1}\left(e^{-\beta\left(\sigma^z_{J-2}\sigma^z_{\ell}+\sigma^x_{J+1}\right)}-\sigma^x_{\ell}\sigma^z_{J-1}\sigma^x_J\sigma^z_{J+1}\right)\nonumber\\
&+\alpha_J\left(e^{-\beta\left(\sigma^z_{J-1}\sigma^x_J+\sigma^x_{J+2}\right)}-\sigma^z_J\sigma^x_{J+1}\sigma^z_{J+2}\right)\nonumber\\
&+\sum_{j\geq J+1}^{L-2}\alpha_j\left(e^{-\beta\left(\sigma^x_j+\sigma^x_{j+2}\right)}-\sigma^z_j\sigma^x_{j+1}\sigma^z_{j+2}\right)\,.
\end{align}
One can verify that the following operators are conserved:
\begin{alignat}{1}
    \bar{\chi}_1 &= \sigma^z_1\,, \\
    \bar{\chi}_2 &=  \sigma^x_1\sigma^x_3\dots\sigma^x_{J-3}\sigma^x_{J-1}\sigma^z_{J}\,,\\
    \bar{\chi}_3&=\sigma^z_\ell\sigma^x_{J+1}\dots\sigma^x_{L-2}\sigma^x_{L}\,,\\ 
    \bar{\chi}_4& = \sigma^x_J\sigma^x_{J+2}\dots\sigma^x_{L-3}\sigma^x_{L-1}\,.
\end{alignat}
They satisfy the nontrivial algebra
\begin{alignat}{2}
    \bar{\chi}_1\bar{\chi}_2&=-\bar{\chi}_2\bar{\chi}_1\,, ~ ~ ~ &\bar{\chi}_2\bar{\chi}_3& = \bar{\chi}_3\bar{\chi}_2\,,\\
    \bar{\chi}_1\bar{\chi}_3&= \bar{\chi}_3\bar{\chi}_1\,, ~ ~ ~ &\bar{\chi}_2\bar{\chi}_4& = -\bar{\chi}_2\bar{\chi}_4\,,\\
    \bar{\chi}_1\bar{\chi}_4&=\bar{\chi}_4\bar{\chi}_1\,, ~ ~ ~ &\bar{\chi}_3\bar{\chi}_4& = \bar{\chi}_4\bar{\chi}_3\,.
\end{alignat}
The smallest irreducible representation of this algebra is two-dimensional
(see Appendix~\ref{app: hg 2-chain} for details), enforcing at least a
twofold degeneracy throughout the spectrum. Indeed, since, e.g., $\bar{\chi}_1$
and $\bar{\chi}_2$ commute with the Hamiltonian but anticommute with each
other, applying $\bar{\chi}_2$ to an energy eigenstate with a definite
$\bar{\chi}_1$ eigenvalue produces an orthogonal eigenstate with the same
energy and the opposite $\bar{\chi}_1$ eigenvalue. In the explicit
half-gauged construction, the QMBS subspace is eight-dimensional, compared
with the four-dimensional subspace of Eq.~\eqref{eq:44}. Its degeneracy
therefore acquires an algebraically protected component, beyond the
degeneracy inherited from the anchor states. The corresponding logical
qubit is described by $\bar{\chi}_1$, localized at the left boundary, and
$\bar{\chi}_2$, a nonlocal string extending from that boundary to the
interface. Half-gauging thus restores a protected-mode interpretation of
the finite-energy degeneracy, with the logical operators now supported
across the boundary and interface. This algebraic protection survives
perturbations that commute with the conserved operators, whereas
preserving the exact QMBS additionally requires maintaining the embedding
condition.

\section{Final remarks \label{secV}}
We have explored the interplay between quantum many-body scars, symmetry defects, and duality interfaces. Focusing on a specific class of models known as $k$-chains, we have shown the conditions under which quantum scarring can persist. The results presented in this paper help illuminate a classification scheme for applying duality transformations to scarred models; specifically, they illustrate when a quantum scarred state can be expected to survive a duality operation, even after the insertion of defects scrambles the original Hilbert space.

We also explored the mechanism of half-gauging and duality interfaces in scarred systems, which may offer a promising route for generating novel symmetry-protected scarred states. Because gauge operations can be related to quantum operations \cite{Khan2024, Tachikawa2024, Verstraete2023, Lootens2025}, they can be implemented using sequential quantum circuits \cite{Xie2023, Vanhove2025, Fontana2025}. A half-gauged system can then be obtained simply by truncating the circuit at a given position. This provides a viable path toward the experimental realization of these systems on near-term digital quantum computers, which are already utilized to explore topological order \cite{satzinger2021realizing} and correlated dynamics \cite{cochran2025visualizing} via sequential circuits.

Looking ahead, our work points to several compelling directions for future inquiry. Because the half-gauging procedure is independent of spatial dimensionality, it opens up the possibility of exploring new QMBS states in higher dimensions, where protection is provided by more exotic, and potentially more robust—symmetries. For instance, the topological protection of quantum many-body scars remains an open question \cite{Chamon2019}, as the mechanisms ensuring protection in the ground state no longer hold for states embedded in the middle of the spectrum. Analyzing the interfaces that arise from half-gauging a p-form symmetry in these scenarios might provide a new perspective on achieving a robust, degenerate subspace for QMBS physics. 

Finally, another natural avenue of investigation concerns the effects of these interfaces on thermalization; determining whether they allow us to surpass existing bounds for the slow thermalization \cite{Motrunich2020} of scarred states remains a fruitful direction for future research.

\section{Acknowledgments}

We thank Masahito Yamazaki for fruitful discussions during the completion of this work. Exact-diagonalization calculations were performed using the QuSpin package \cite{Quspin1,Quspin2} in Python. The code and numerical data used in this work are publicly available on Zenodo~\cite{FontanaOlivieroHuang2026Code}.
W.B.F. is supported by the Theory of Quantum Matter Unit at the Okinawa Institute of Science and Technology (OIST). This work was initiated while W.B.F. was supported through the MOST Young Scholar Fellowship awarded to Y.-P.H. (Grant No.~114-2636-M-007-001-).
F.G.O. acknowledges financial support from the National Center for Theoretical Sciences under Grant No.~115-2124-M-002-014-.
Y.-P.H. acknowledges support from the National Science and Technology Council, Taiwan, through an individual research grant (Grant No.~115-2112-M-007-049-), as well as support from the National Center for Theoretical Sciences (Grant No.~115-2124-M-002-014-) and from the Yushan Young Scholar Program of the Ministry of Education, Taiwan, through an Administrative Support Grant.

ChatGPT 5.6 was used to assist with code development and language editing.

\bibliographystyle{ref_style.bst}

\bibliography{sptref}
\appendix
\section*{Appendices}
We reserve this appendices to discuss the half-gauging procedure of a few other $k$-chains, to show that indeed, there is a pattern that increases the dimensionality of the QMBS subspace, in particular, it always doubles. The size of the protected subspace (the irreducible representation realized by the mixed symmetries in the half-gauged chain) will depend on the specific chain. 

We also reserve this appendices to display some interesting results regarding revivals of a perturbed $1$-chain. Surprinsingly, the $1$-chain in the absence of symmetry defects, display perfect revivals when perturbed with a symmetry breaking term, given that the strength of the perturbation is not large enough. This suggests that these scarred $k$-chains, may retain their non-thermalizing properties even in the presence of symmetry breaking terms that would, in principle, lead to fast thermalization. 

\section{Further results on the half-gauging of larger scarred $k$-chains \label{app1}}
In this appendix, we perform the half-gauging procedure on the $2$- and $3$-chains, to show the explicit emergence of a zero mode emanating from the interface, that unavoidably lead to the doubling on the size of the underlying scarred subspace.

\subsection{Half-Gauging scarred 2-chain \label{app: hg 2-chain}}
In this section we want to analyze the half-gauging on the scarred 2-chain with open boundary conditions (OBC). The model is intially defined as
\begin{equation}
    H^{(2)}_{s}=\sum_{j=1}^{L-2}\alpha_j\left(e^{-\beta\left(X_j+X_{j+2}\right)}-Z_jX_{j+1}Z_{j+2}\right)\,.
    \label{eq:A9}
\end{equation}

Before we proceed, it is interesting to notice a few features of the model. notice that for $\beta=0$, the model \eqref{eq:A9} becomes simply the standard cluster model with OBC, a famous stabilizer model with protected symmetry protected edge modes\cite{Kitaev2011, Cirac2011, Xie2011, Nayak2014,Sarma2015, Alicea2016}. The edge modes can be seen from the quantities at each edge
\begin{alignat}{2}
    \chi_1&=X_1Z_2\,, ~ ~ ~ &\chi_2& = Z_1\,,\\
    \zeta_1&=Z_{L-1}X_L\,, ~ ~ ~ &\zeta_2& = Z_L\,,
\end{alignat}
that commute with the Hamiltonian $H^{(2)}_s(0)$, but anticommute with eachother, i.e., $\{\chi_1,\,\chi_2\}=\{\zeta_1,\,\zeta_2\}=0$. These 2 qubits at each edge are responsible for the 4-fold topological degeneracy of the model. When $\beta\neq0$, a 4-fold space will still be embbeded at the middle of the spectrum (due to the missing cluster stabilizers with OBC), but the edge modes are no longer conserved quantities. Therefore, the scarred subspace still has a 4-fold degeneracy but do not necessarily keep some the properties important for quantum computation \cite{Sarma2015, Alicea2016}. We now proceed to half-gauge the model to show that, even with its scarred version, we can recover a notion of a protected qubit, except that it will no longer be localized at the edge, but instead, it can be thought as a quantity emanating from the interface induced by the half-gauging.

As usual, the model has a $\mathbb{Z}_2\times\mathbb{Z}_2$ global symmetry that we want to promote to a local one through gauging on only half of the chain. This can be done, as in the main text, by inserting extra degrees of freedom up to some arbitrary site $J$. In the current scenario, we have to be mindful of the boundary terms. For that, consider the gauged model as 
\begin{align}
H^{(2)}_{sG}&=\alpha_1\left(e^{-\beta\left(X_1+X_3\right)}-Z_1X_2\tau^x_2Z_3\right)\nonumber\\
&+\sum_{j=2}^{J-2}\alpha_j\left(e^{-\beta\left(X_j+X_{j+2}\right)}-Z_jX_{j+1}\tau^x_{j+1}Z_{j+2}\right)\nonumber\\
&+\alpha_{J-2}\left(e^{-\beta\left(X_{J-2}+X_{J}\right)}-Z_{J-2}X_{J-1}\tau^x_{J-1}Z_{J}\right)\nonumber\\
&+\alpha_{J-1}\left(e^{-\beta\left(X_{J-1}+X_{J+1}\right)}-Z_{J-1}X_J\tau^x_{J}Z_{J+1}\right)\nonumber\\
&+\sum_{j\geq J}^{L-2}\alpha_j\left(e^{-\beta\left(X_j+X_{j+2}\right)}-Z_jX_{j+1}Z_{j+2}\right)\,.
\end{align}

The above model is invariant under the set of local Gauss laws
\begin{align}
    G_1&=X_1\tau^z_2\,,\\
    G_j&=\tau^z_{j-1}X_j\tau^z_{j+1}\,, ~ ~ ~ j=2,\,\dots,\,J-1\,,
\end{align}
as before, notice that even though the Hilbert space has been extended by a factor of $2^J$, we end up with only $2^{J-1}$ constraints. Furthermore, due to OBC, the Gauss law constraint at the boundary has to be modified. As usual, we now impose the constraints $G_j=1$ for all $j$, and we obtain
\begin{align}
&H^{(2)}_{sG}=\alpha_1\left(e^{-\beta\left(\tau^z_2+\tau^z_{2}\tau^z_4\right)}-\tau^z_1\tau^z_3Z_1\tau^x_2Z_3\right)\nonumber\\
&+\sum_{j=2}^{J-2}\alpha_j\left(e^{-\beta\left(\tau^z_{j-1}\tau^z_{j+1}+\tau^z_{j+1}\tau^z_{j+3}\right)}-\tau^z_{j}\tau^z_{j+2}Z_j\tau^x_{j+1}Z_{j+2}\right)\nonumber\\
&+\alpha_{J-2}\left(e^{-\beta\left(\tau^z_{J-3}\tau^z_{J-1}+X_{J}\right)}-\tau^z_{J-2}\tau^z_{J}Z_{J-2}\tau^x_{J-1}Z_{J}\right)\nonumber\\
&+\alpha_{J-1}\left(e^{-\beta\left(\tau^z_{J-2}\tau^z_J+X_{J+1}\right)}-Z_{J-1}\tau^x_{J}X_JZ_{J+1}\right)\nonumber\\
&+\sum_{j\geq J}^{L-2}\alpha_j\left(e^{-\beta\left(X_j+X_{j+2}\right)}-Z_jX_{j+1}Z_{j+2}\right)\,.
\end{align}

Define new spin variables that are gauge invariant through
\begin{alignat}{2}
    \hat{Z}_j &=\tau_j^z ,\,~ ~ ~ &j&=1,\,\dots,\,J\,,\\
    \hat{X}_{j+1} &=Z_j\tau^x_{j+1}Z_{j+2} \,, ~ ~ ~ &j&=2,\,\dots,\,J-1\,,\\
    \hat{X}_{J-1} &=Z_{J-2}\tau^x_{J-1}\,,\\
    \hat{X}_J &=Z_{J-1}\tau^x_J\,,
\end{alignat}
then we get
\begin{align}
H^{(2)}_{sG}&=\alpha_1\left(e^{-\beta\left(\hat{Z}_2+\hat{Z}_{2}\hat{Z}_4\right)}-\hat{Z}_1\hat{X}_2\hat{Z}_3\right)\nonumber\\
&+\sum_{j=2}^{J-2}\alpha_j\left(e^{-\beta\left(\hat{Z}_{j-1}\hat{Z}_{j+1}+\hat{Z}_{j+1}\hat{Z}_{j+3}\right)}-\hat{Z}_{j}\hat{X}_{j+1}\hat{Z}_{j+2}\right)\nonumber\\
&+\alpha_{J-2}\left(e^{-\beta\left(\hat{Z}_{J-3}\hat{Z}_{J-1}+X_{J}\right)}-\hat{Z}_{J-2}\hat{X}_{J-1}\hat{Z}_{J}Z_{J}\right)\nonumber\\
&+\alpha_{J-1}\left(e^{-\beta\left(\hat{Z}_{J-2}\hat{Z}_J+X_{J+1}\right)}-\hat{X}_{J}X_JZ_{J+1}\right)\nonumber\\
&+\sum_{j\geq J}^{L-2}\alpha_j\left(e^{-\beta\left(X_j+X_{j+2}\right)}-Z_jX_{j+1}Z_{j+2}\right)\,.
\label{eq:A20}
\end{align}

As we had in the main text, the Hamiltonian \eqref{eq:A20} can be written using a single set of spin operators at the cost of introducing an extra site $\ell$ between the sites $J-1$ and $J$ in our chain. Thus, we can define a new set of variables
\begin{alignat}{2}
    \hat{Z}_j &= \sigma^z_j\,, ~ ~ ~ j&=&2,\,\dots,\,J-1\,,\\
    \hat{X}_j &= \sigma^x_j\,, ~ ~ ~ j&=&2,\,\dots,\,J-1\,,\\
    \hat{Z}_J&=\sigma^z_\ell\,, \\
    \hat{X}_J&=\sigma^x_\ell\,,\\
    X_j &= \sigma^x_j\,, ~ ~ ~ j&=&J+1,\,\dots,\,L\,,\\
    Z_j &= \sigma^z_j\,, ~ ~ ~ j&=&J+1,\,\dots,\,L\,,
\end{alignat}

which allow us to write down our Hamiltonian \eqref{eq:A20} as
\begin{align}
H^{(2)}_{sG}&=\alpha_1\left(e^{-\beta\left(\sigma^z_2+\sigma^z_{2}\sigma^z_4\right)}-\sigma^z_1\sigma^x_2\sigma^z_3\right)\nonumber\\
&+\sum_{j=2}^{J-2}\alpha_j\left(e^{-\beta\left(\sigma^z_{j-1}\sigma^z_{j+1}+\sigma^z_{j+1}\sigma^z_{j+3}\right)}-\sigma^z_{j}\sigma^x_{j+1}\sigma^z_{j+2}\right)\nonumber\\
&+\alpha_{J-2}\left(e^{-\beta\left(\sigma^z_{J-3}\sigma^z_{J-1}+\sigma^x_{J}\right)}-\sigma^z_{J-2}\sigma^x_{J-1}\sigma^z_{\ell}\sigma^z_{J}\right)\nonumber\\
&+\alpha_{J-1}\left(e^{-\beta\left(\sigma^z_{J-2}\sigma^z_{\ell}+\sigma^x_{J+1}\right)}-\sigma^x_{\ell}\sigma^x_J\sigma^z_{J+1}\right)\nonumber\\
&+\sum_{j\geq J}^{L-2}\alpha_j\left(e^{-\beta\left(\sigma^x_j+\sigma^x_{j+2}\right)}-\sigma^z_j\sigma^x_{j+1}\sigma^z_{j+2}\right)\,.
\label{eq:A27}
\end{align}

As a final step, it is also convenient to conjugate the Hamiltonian with a local $CZ_{J-1,\,J}$ gate, such that we can write our Hamiltonian as
\begin{align}
H^{(2)}_{sG}&=\alpha_1\left(e^{-\beta\left(\sigma^z_2+\sigma^z_{2}\sigma^z_4\right)}-\sigma^z_1\sigma^x_2\sigma^z_3\right)\nonumber\\
&+\sum_{j=2}^{J-2}\alpha_j\left(e^{-\beta\left(\sigma^z_{j-1}\sigma^z_{j+1}+\sigma^z_{j+1}\sigma^z_{j+3}\right)}-\sigma^z_{j}\sigma^x_{j+1}\sigma^z_{j+2}\right)\nonumber\\
&+\alpha_{J-2}\left(e^{-\beta\left(\sigma^z_{J-3}\sigma^z_{J-1}+\sigma^z_{J-1}\sigma^x_{J}\right)}-\sigma^z_{J-2}\sigma^x_{J-1}\sigma^z_{\ell}\right)\nonumber\\
&+\alpha_{J-1}\left(e^{-\beta\left(\sigma^z_{J-2}\sigma^z_{\ell}+\sigma^x_{J+1}\right)}-\sigma^x_{\ell}\sigma^z_{J-1}\sigma^x_J\sigma^z_{J+1}\right)\nonumber\\
&+\alpha_J\left(e^{-\beta\left(\sigma^z_{J-1}\sigma^x_J+\sigma^x_{J+2}\right)}-\sigma^z_J\sigma^x_{J+1}\sigma^z_{J+2}\right)\nonumber\\
&+\sum_{j\geq J+1}^{L-2}\alpha_j\left(e^{-\beta\left(\sigma^x_j+\sigma^x_{j+2}\right)}-\sigma^z_j\sigma^x_{j+1}\sigma^z_{j+2}\right)\,.
\label{eq:A28}
\end{align}

At this point, we can examine the symmetry content of the Hamiltonian \eqref{eq:A28}. It is straightforward to check that the following quantities are conserved (i.e., commute with $H^{(2)}_{sG}$):
\begin{alignat}{1}
    \bar{\chi}_1 &= \sigma^z_1\,, \\
    \bar{\chi}_2 &=  \sigma^x_1\sigma^x_3\dots\sigma^x_{J-3}\sigma^x_{J-1}\sigma^z_{J}\,,\\
    \bar{\chi}_3&=\sigma^z_\ell\sigma^x_{J+1}\dots\sigma^x_{L-2}\sigma^x_{L}\,,\\ 
    \bar{\chi}_4& = \sigma^x_J\sigma^x_{J+2}\dots\sigma^x_{L-3}\sigma^x_{L-1}\,.
\end{alignat}

These set of symmetry operators $\bar{\chi}_i$ obey an interesting algebra given as follows
\begin{alignat}{2}
    \bar{\chi}_1\bar{\chi}_2&=-\bar{\chi}_2\bar{\chi}_1\,, ~ ~ ~ &\bar{\chi}_2\bar{\chi}_3& = \bar{\chi}_3\bar{\chi}_2\,,\\
    \bar{\chi}_1\bar{\chi}_3&= \bar{\chi}_3\bar{\chi}_1\,, ~ ~ ~ &\bar{\chi}_2\bar{\chi}_4& = -\bar{\chi}_4\bar{\chi}_2\,,\\
    \bar{\chi}_1\bar{\chi}_4&=\bar{\chi}_4\bar{\chi}_1\,, ~ ~ ~ &\bar{\chi}_3\bar{\chi}_4& = \bar{\chi}_4\bar{\chi}_3\,.
\end{alignat}

The dimensionality of this algebra can be obtained by noticing that a good basis is given by the eigenstates of $\bar{\chi}_1,\,\bar{\chi}_3,\,\bar{\chi}_4$, since these operators commute with the Hamiltonian and with one another. Then, we notice that the effect of the $\bar{\chi}_2$ operator is to give us a new set of eigenstates with flipped eigenvalues for $\bar{\chi}_1$ and $\bar{\chi}_4$, i.e., 
\begin{align}
    \bar{\chi}_1\left(\bar{\chi}_2\ket{\bar{\chi}_1,\,\bar{\chi}_3,\,\bar{\chi}_4}\right) = -x_1\left(\bar{\chi}_2\ket{\bar{\chi}_1,\,\bar{\chi}_3,\,\bar{\chi}_4}\right)\,,\\
\bar{\chi}_4\left(\bar{\chi}_2\ket{\bar{\chi}_1,\,\bar{\chi}_3,\,\bar{\chi}_4}\right) = -x_4\left(\bar{\chi}_2\ket{\bar{\chi}_1,\,\bar{\chi}_3,\,\bar{\chi}_4}\right)\,,
\label{eq:logical operators}
\end{align}
where $x_1,\,x_4$ are the corresponding eigenvalues associated to $\bar{\chi}_1,\,\bar{\chi}_4$. Thus, by a simple counting argument, we can notice that the new degeneracy is 8-fold, doubling from the initial 4-fold degeneracy we had at first. The extra 4-fold degeneracy stems from the new zero modes emanating from the topological interface. Another, perhaps simpler, way of understanding this degeneracy is computing the rank of the commutation matrix

\begin{equation}
    A_{ij} = \begin{cases}
        1,\, ~ ~ ~ \text{if} ~ ~ ~ \{\bar{\chi}_i,\,\bar{\chi}_j\}=0\,,\\
        0,\,~ ~ ~ \text{if} ~ ~ ~~ \,  [\bar{\chi}_i,\,\bar{\chi}_j]=0\,.
    \end{cases}
    \label{eq: comm matrix}
\end{equation}

Half of the rank of $A_{ij}$ give us the size of the irreducible representation generated by the algebra of $\{\bar{\chi}_i\}$ (equivalently, the number of logical qubits encoded is half the rank of the matrix $A_{ij}$) \cite{Gottesman1997, NielsenChuang2010}.It is straightforward to check that the rank of the $A$-matrix for the symmetry operators \eqref{eq:logical operators} is $r=2$, which tell us that the irreps of this algebra are $2^{r/2}=2$-dimensional. Meanwhile, the size of the scarred subspace is $\dim(\mathcal{S})=2^{n-r/2}=2^{4-1}=8$, with $n$ the number of symmetry generators in \eqref{eq:logical operators}, which agrees with our previous analysis. Thus, half-gauging the scarred $2$-chain, gave us a protected $2$-fold degeneracy (an actual encoded qubit) effectively doubling the original subspace.

\subsection{Half-Gauging scarred $3$-chain}

For the scarred $3$-chain the half-gauging with OBC follows a similar procedure. We start with basic model
\begin{equation}
    H^{(3)}_{s} = \sum_j^{L-3}\alpha_j\left(e^{-\beta(X_j+X_{j+1})} - Z_jX_{j+1}X_{j+2}Z_{j+3}\right)\,.
    \label{eq:A30}
\end{equation}

The goal is to half gauge the $\mathbb{Z}_2\times\mathbb{Z}_2\times\mathbb{Z}_2$ global symmetry of the model. As before, the initial degeneracy of the model can already be seen from the three missing stabilizers in the Hamiltonian \eqref{eq:A30}, which lead to a scarred subspace of dimension $2^3=8$. As before, after insertion of the topological interface, this subspace will increase in size, which can be attributed to the existence the non local zero modes emanating from the interface.

We want the gauged model to be invariant under the set of Gauss laws $G_j = \tau^z_{j-1}X_j\tau^z_{j+2}$, with the appropriate modifications required at the left boundary. With that, we consider that the half-gauged Hamiltonian is written as 
\begin{align}
    &H^{(3)}_{sG} = \alpha_1\left(e^{-\beta(X_1+X_4)} -Z_1X_2X_3\tau^x_3Z_4\right)\nonumber\\
    &+\alpha_2\left(e^{-\beta(X_2+X_5)} -Z_2X_3X_4\tau^x_4Z_5\right)\nonumber\\
    &+\sum_{j=3}^{J-3}\alpha_j\left(e^{-\beta(X_j+X_{j+3})} -Z_jX_{j+1}X_{j+2}\tau^x_{j+2}Z_{j+3}\right)\nonumber\\
    &+\alpha_{J-2}\left(e^{-\beta(X_{J-2}+X_{J+1})} -Z_{J-2}X_{J-1}X_J\tau^x_JZ_{J+1}\right)\nonumber\\
    &+\sum_{j\geq J-1}^{L-3}\alpha_j\left(e^{-\beta(X_j+X_{j+3})} -Z_jX_{j+1}X_{j+2}Z_{j+3}\right)\,.
    \label{eq:A31}
\end{align}

As expected, the Hamiltonian \eqref{eq:A31} was constructed such that it is invariant under the local set of conserved quantities
\begin{align}
    G_1 &= X_1\tau^z_{3}\,,\\
    G_j& = \tau^z_{j-1}X_j\tau^z_{j+2}\,, ~ ~ ~ j=2,\,\dots,\,J-2\,.
    \label{eq: gauss law constraints 3-chain}
\end{align}
We call attention to the fact that, with the $3$-chain, the number of Gauss law constraints is reduced by one compared to the $2$-chain. This is a consequence of the gauge prescription, once we adapt it to the larger separation between the spin operators. After we project the gauged model \eqref{eq:A31} into the physical Hilbert space and write it in terms of the gauge invariant quantities
\begin{alignat}{2}
    \hat{Z}_j &= \tau^z_j,\,~ ~ ~ &j=&\,1,\,\dots,\,J\,,\\
    \hat{X}_j &= Z_{j-1}\tau^x_{j}Z_{j+2}\,,~ ~ ~ &j=&\,3,\,\dots,\,J-3\,,\\
    \hat{X}_{J-2} &= Z_{J-4}\tau^x_{J-2}\,,\\
    \hat{X}_{J-1} &= Z_{J-3}\tau^x_{J-1}\,,\\
    \hat{X}_{J} &= Z_{J-2}\tau^x_J\,,
\end{alignat}
we get the following model
\begin{widetext}
\begin{align}
    &H^{(3)}_{sG} = \alpha_1\left(e^{-\beta\left(\hat{Z}_3+\hat{Z}_3\hat{Z}_6\right)} - \hat{Z}_1\hat{Z}_2\hat{X}_3\hat{Z}_4\hat{Z}_5\right)
    +\sum_j\alpha_j\left(e^{-\beta\left(\hat{Z}_{j-1}\hat{Z}_{j+2}+\hat{Z}_{j+2}\hat{Z}_{j+5}\right)} - \hat{Z}_j\hat{Z}_{j+1}\hat{X}_{j+2}\hat{Z}_{j+3}\hat{Z}_{j+4}\right)\nonumber\\
    &+\alpha_{J-4}\left(e^{-\beta\left(\hat{Z}_{J-5}\hat{Z}_{J-2}+X_{J-1}\right)} - \hat{Z}_{J-4}\hat{Z}_{J-3}\hat{X}_{J-2}\hat{Z}_{J-1}\hat{Z}_{J}Z_{J-1}\right)
    +\alpha_{J-3}\left(e^{-\beta\left(\hat{Z}_{J-4}\hat{Z}_{J-1}+X_{J}\right)} - \hat{Z}_{J-3}\hat{Z}_{J}\hat{X}_{J-1}X_{J-1}Z_{J}\right)\nonumber\\
    &+\alpha_{J-2}\left(e^{-\beta\left(\hat{Z}_{J-3}\hat{Z}_{J}+X_{J+1}\right)} -\hat{X}_{J}X_{J-1}X_{J}Z_{J+1}\right)
    +\sum_{j\geq J-1}^{L-3}\alpha_j\left(e^{-\beta\left(X_j+X_{j+3}\right)} - Z_jX_{j+1}X_{j+2}Z_{j+3}\right)\,.
\end{align}
\end{widetext}

In order to put this Hamiltonian in a more appealing form, we first write it in terms of a single set of Pauli operators. To achieve that, we must first notice that we have to increase our chain by two sites. This is a consequence of having $J-2$ Gauss law constraints in \eqref{eq: gauss law constraints 3-chain}, instead of $J$. We label these two extra sites by $\ell_1$ and $\ell_2$ and they are inserted between the original $J-2$ and $J-1$ sites. Therefore, the ordering of sites in the new chain is $(\dots,\, J-2,\, \ell_1,\,\ell_2,\,J-1,\, J,\,\dots)$. The new variables are then defined as
\begin{alignat}{2}
    \sigma^z_j &= \hat{Z}_j\,, ~ ~ ~ &j=&\,1,\,\dots,\,J-2,\,J+1,\,\dots,\,L\,,\\
    \sigma^x_j &= \hat{X}_j\,, ~ ~ ~ &j=&\,1,\,\dots,\,J-2,\,J+1,\,\dots,\,L\,,\\
    \sigma^z_{\ell_1} &= \hat{Z}_{J-1}\,,\\
    \sigma^z_{\ell_2} &= \hat{Z}_{J}\,,\\
    \sigma^x_{\ell_1} &= \hat{X}_{J-1}\,,\\
    \sigma^x_{\ell_x} &= \hat{Z}_{J}\,.
\end{alignat}

After that, we also consider the action of two controlled-Z operations, $CZ_{\ell_1,\,J}$ and $CZ_{\ell_2,\,J+1}$, in terms of the new spin variables $\Vec{\sigma}$. With that, we reach the final form of half-gauged Hamiltonian
\begin{widetext}
\begin{align}
    &H^{(3)}_{sG} = \alpha_1\left(e^{-\beta\left(\sigma^z_3+\sigma^z_3\sigma^z_6\right)} - \sigma^z_1\sigma^z_2\sigma^x_3\sigma^z_4\sigma^z_5\right)
    +\sum_j\alpha_j\left(e^{-\beta\left(\sigma^z_{j-1}\sigma^z_{j+2}+\sigma^z_{j+2}\sigma^z_{j+5}\right)} - \sigma^z_j\sigma^z_{j+1}\sigma^x_{j+2}\sigma^z_{j+3}\sigma^z_{j+4}\right)\nonumber\\
    &+\alpha_{J-4}\left(e^{-\beta\left(\sigma^z_{J-5}\sigma^z_{J-2}+\sigma^z_{J-2}\sigma^x_{J-1}\right)} - \sigma^z_{J-4}\sigma^z_{J-3}\sigma^x_{J-2}\sigma^z_{\ell_1}\sigma^z_{\ell_2}\right)
    +\alpha_{J-3}\left(e^{-\beta\left(\sigma^z_{J-4}\sigma^z_{\ell_1}+\sigma^z_{\ell_1}\sigma^x_{J}\right)} - \sigma^z_{J-3}\sigma^z_{J-2}\sigma^x_{\ell_1}\sigma^z_{\ell_2}\sigma^x_{J-1}\right)\nonumber\\
    &+\alpha_{J-2}\left(e^{-\beta\left(\sigma^z_{J-3}\sigma^z_{\ell_2}+\sigma^x_{J+1}\right)} -\sigma^z_{\ell_1}\sigma^x_{\ell_2}\sigma^x_{J-1}\sigma^x_{J}\sigma^z_{J+1}\right)
    +\alpha_{J-1}\left(e^{-\beta\left(\sigma^z_{J-2}\sigma^x_{J-1}+\sigma^x_{J+2}\right)} - \sigma^z_{J-1}\sigma^z_{\ell_1}\sigma^x_{J}\sigma^x_{J+1}\sigma^z_{J+2}\right)\nonumber\\
    &+\alpha_{J}\left(e^{-\beta\left(\sigma^z_{\ell_1}\sigma^x_{J}+\sigma^x_{J+3}\right)} - \sigma^z_J\sigma^x_{J+1}\sigma^x_{J+2}\sigma^z_{J+3}\right)
   +\sum_{j\geq J}^{L-3}\alpha_j\left(e^{-\beta\left(\sigma^x_j+\sigma^x_{j+3}\right)} - \sigma^z_j\sigma^x_{j+1}\sigma^x_{j+2}\sigma^z_{j+3}\right)\,,
    \label{eq:A46}
\end{align}
\end{widetext}
where in \eqref{eq:A46} we wrote the Hamiltonian such to make explicit the spins at the corresponding new sites. As before, the half-gauged Hamiltonian has a interesting set of conserved quantities that form a non-trivial algebra due to the existence of the interface. For example, the set of symmetries for the Hamiltonian \eqref{eq:A46} is given as 
\begin{align}
    \bar{\zeta}_1 &=\sigma^z_1\,\\
    \bar{\zeta}_2 &=\sigma^z_2\,\\
    \bar{\zeta}_3 &=\sigma^x_1\sigma^x_4\dots\sigma^x_{\ell_1}\sigma^z_J\,\\
    \bar{\zeta}_4 &=\sigma^x_2\sigma^x_5\dots\sigma^x_{J-2}\sigma^z_{J-1}\,\\
    \bar{\zeta}_5 &=\sigma^x_J\sigma^x_{J+3}\dots\sigma^x_{L}\,\\
    \bar{\zeta}_6 &=\sigma^z_{\ell_2}\sigma^x_{J+1}\sigma^x_{J+4}\dots\sigma^x_{L-1}\,\\
    \bar{\zeta}_7 &=\sigma^x_{J-1}\sigma^x_{J+2}\dots\sigma^x_{L-2}\,.
    \label{eq: 3-chain symm}
\end{align}

The algebra generated by the set of symmetries $\bar{\zeta}_i$ is a bit more complicated than it was for the $2$-chain, but it is still manageable. In particular, the only operators $\bar{\zeta}_i$ that anticommute can be written as 
\begin{alignat}{1}
    \bar{\zeta}_1\bar{\zeta}_3&=-\bar{\zeta}_3\bar{\zeta}_1\,,\\
    \bar{\zeta}_2\bar{\zeta}_4&=-\bar{\zeta}_4\bar{\zeta}_2\,,\\
    \bar{\zeta}_3\bar{\zeta}_5&=-\bar{\zeta}_5\bar{\zeta}_3\,,
\end{alignat}
with all others having a trivial commutator. 

To obtain the size of the representation of the algebra generated by $\{\bar{\zeta_i}\}$, we can once again write down the commutation $A$-matrix \eqref{eq: comm matrix} for the set of symmetries \eqref{eq: 3-chain symm}. Specifically, the matrix assumes 
\begin{equation}
    A=\begin{pmatrix}
        0&0&1&0&0&0&0\\
        0&0&0&1&0&0&0\\
        1&0&0&0&1&0&0\\
        0&1&0&0&0&0&0\\
        0&0&1&0&0&0&0\\
        0&0&0&0&0&0&0\\
        0&0&0&0&0&0&0
    \end{pmatrix}\,.
\end{equation}
\vspace{1pt}

We can observe that the above matrix has rank $r=4$, which tell us that the irreps of the algebra \eqref{eq: 3-chain symm} are $2^{4/2}=4$-dimensional. Therefore, the scarred subspace of half-gauged $3$-chain encodes two qubits. The total degeneracy of the subspace follows from $2^{n-r/2}=2^{5}=32$. Which is in accordance to the fact that, the original (pre-half gauging) subspace had dimension $8$, and after encoding two qubits we get an $8\times4=32$-dimensional subspace.

\section{Coherent two-level dynamics from the scar doublet in perturbed 1-chain}

Consider the simple chain with $k=1$ in Eq. (\ref{eq:scarred models}) in the absence of symmetry defects
\begin{equation}
H^{(1)}_s = \sum_{j=0}^{L-1} \alpha_j\left(e^{-\beta\left(X_j+X_{j+1}\right)} -Z_jZ_{j+1}\right)\,.
\end{equation}
As we discussed in the main text, we can construct a $\mathbb{Z}_2$ degenerate scarred subspace spanned by $|E_{\pm}\rangle=e^{\frac{\beta}{2}\sum X_j}|\psi_{\pm}\rangle$, such that $H^{(1)}_s |E_{\pm}\rangle=0$. The scars states are the symmetric/antisymmetric linear combinations $\mathcal{S_{\pm}}=\frac{1}{\sqrt{2}}\left(|E_{+}\rangle\pm|E_{-}\rangle\right)$.
We investigate the real-time dynamics of the symmetry-resolved zero-energy scar states,
$|\mathcal{S}_+\rangle$ and $|\mathcal{S}_-\rangle$, after introducing the symmetry breaking perturbation $H(h_z)=H_s ^{1}+h_z\sum_j Z_j$. For that matter we use the Loschmidt echo,
\begin{equation}
\mathcal{L}(t)=\left|\langle\psi_a(0)|\psi_a(t)\rangle\right|^2 ,
\end{equation}
where $|\psi(t)\rangle=\exp[-iH(h_z)t]|\mathcal{S_{\pm}}\rangle$.

Interestingly, the dynamics display pronounced revivals, which are not associated with a conventional many-body scar tower of scars \cite{Mougdalya2020}. To understand the origin of these oscillations, we analyze the spectral decomposition of the initial state in the eigenbasis of the perturbed Hamiltonian,
\begin{equation}
|\psi_a\rangle=\sum_n c_n |E_n\rangle .
\end{equation}

The spectral weights reveal that the state has support almost exclusively on two eigenstates:
\begin{equation}
|c_1|^2\simeq 0.4993,\qquad |c_2|^2\simeq 0.4991 ,
\end{equation}
with energies
\begin{equation}
E_1=-0.4448,\qquad E_2=0.4417 .
\end{equation}
The remaining spectral weight is approximately $1.5\times10^{-3}$.

Furthermore, the projection onto this two-dimensional subspace,
\begin{equation}
P(t)=
\sum_{n=1,2}
|\langle E_n|\psi_a(t)\rangle|^2 ,
\end{equation}
remains constant during the evolution,
\begin{equation}
P(t)=0.9984883155 ,
\end{equation}
showing that the dynamics are confined to an almost perfect two-level subspace.
\begin{figure}
\centering
\includegraphics[width=1.0\linewidth]{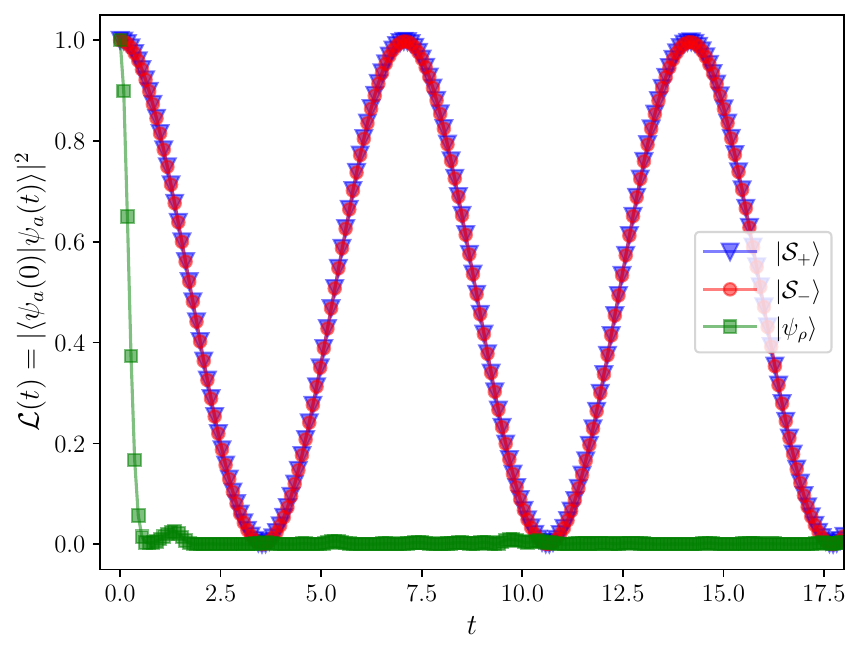}
\caption{\justifying
Loschmidt echo of the scar states compared with a generic high-energy volume-law state. 
The strong revivals of $|\mathcal{S}_\pm\rangle$ originate from coherent oscillations within an almost isolated two-level subspace, while a random state in the full Hilbert space state (in green) rapidly loses memory of its initial condition. Exact diagonalization of a chain with $L=10,\, \alpha=0.3,\, \beta=0.3,\, h_z= 0.05$.}
\label{fig:Loschmidt_1_chain}
\end{figure}
The Loschmidt echo can therefore be understood as coherent oscillations between the two dominant eigenstates,
\begin{equation}
\mathcal{L}\approx
\left|
\frac{1}{2}e^{-iE_1t}
+
\frac{1}{2}e^{-iE_2t}
\right|^2 ,
\end{equation}
which gives
\begin{equation}
\mathcal{L}\approx
\cos^2\left(\frac{\Delta E}{2}t\right),
\end{equation}
where
\begin{equation}
\Delta E=E_2-E_1\simeq0.8865 .
\end{equation}
The corresponding revival period,
\begin{equation}
T=2\pi/\Delta E\simeq7.09 ,
\end{equation}
agrees with the numerical revivals as shown in Fig \ref{fig:Loschmidt_1_chain}.
The origin of this behavior can be understood from the original zero-energy scar doublet. The unperturbed states $|E_+\rangle$ and $|E_-\rangle$ belong to opposite $\mathbb{Z}_2$ sectors. The perturbation mixes these states, $\langle E_+|h_z\sum_j Z_j|E_-\rangle\neq0$, producing two new eigenstates that are approximately equal superpositions of the original scar states. Consequently, the dynamics are effectively described by a two-level Hamiltonian,
\begin{equation}
H_{\rm eff}
=
\begin{pmatrix}
0 & \Delta \\
\Delta & 0
\end{pmatrix}.
\end{equation}
leading to Rabi-like oscillations.
This mechanism differs from the standard many-body scar scenario based on an equally spaced tower of scarred eigenstates. Instead, the observed revivals originate from the coherent dynamics of an isolated scar doublet.

\end{document}